# The evolution of AI from image interpretation toward scientific inference in nanoparticle electron microscopy

*Evropi Toulkeridou[1]*, Jiafei Li[2], Leonardo Lari[3,4], and Panagiotis Grammatikopoulos[5]**

[1]Statistical Analysis and Optimal Design of Experiments Group, Department of Mathematics, University of Castilla-La Mancha, 45071 Toledo, Spain

[2]Materials Science and Engineering, Guangdong Technion – Israel Institute of Technology, Shantou, Guangdong 515063, China

[3]LAME Laboratory, Area Science Park Campus di Basovizza, SS14, KM163.5, 34149, Trieste, Italy

[4]The University of York, School of Physics, Electronics and Technology, Heslignton, YO10 5DD, York, UK

[5]Regional Institute for Applied Scientific Research (IRICA) and Department of Physics, University of Castilla-La Mancha, 13071, Ciudad Real, Spain

Email: evropi.toulkeridou@uclm.es

Email: p.grammatikopoulos@uclm.es

## ABSTRACT

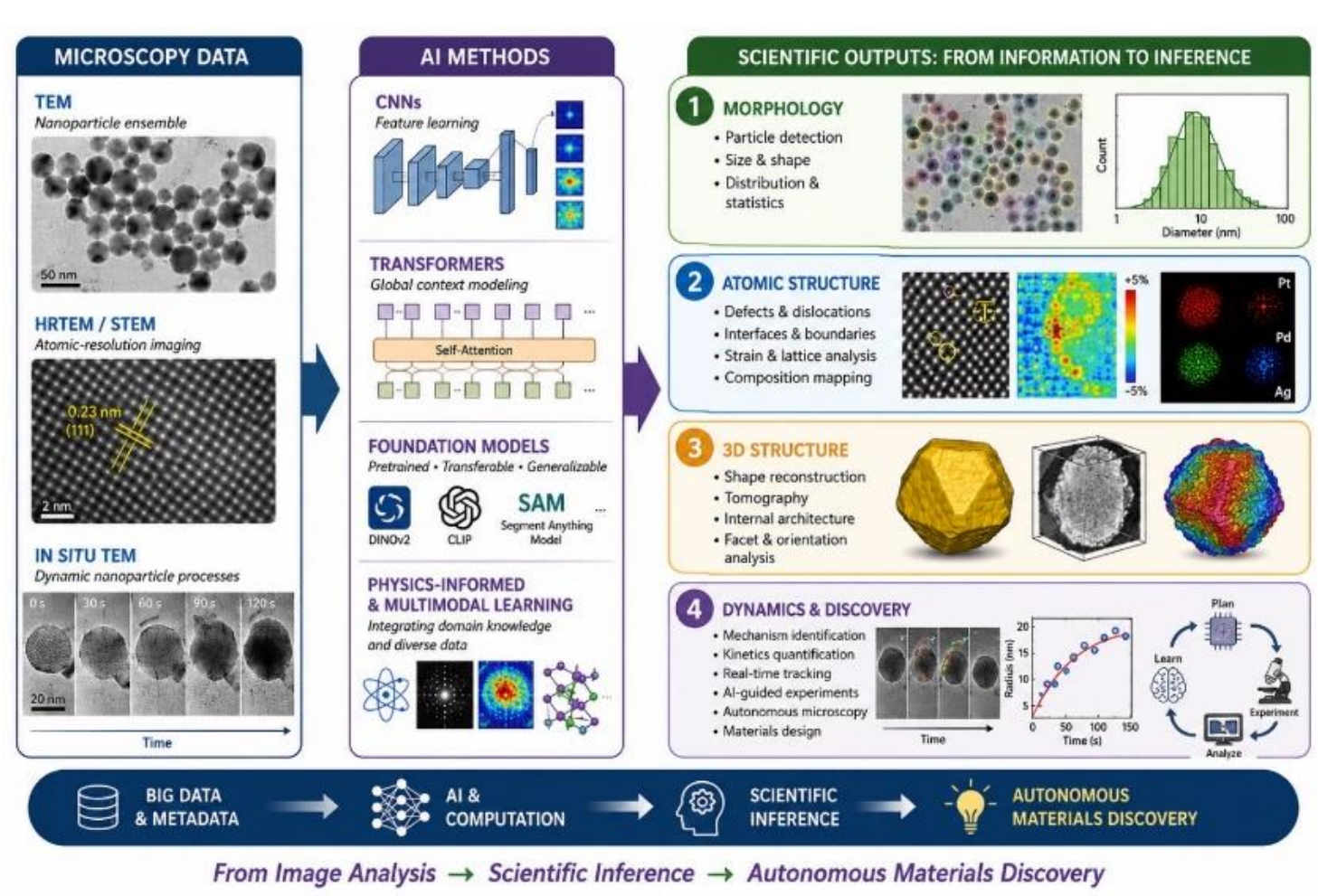

Artificial intelligence (AI) is transforming electron microscopy by enabling quantitative analysis of increasingly large and complex datasets for nanoparticle characterization. Recent advances in machine learning (ML) and deep learning (DL) have expanded microscopy from a descriptive imaging technique into a data-driven platform for structural interpretation, dynamic analysis, and scientific inference. This review examines AI methodologies for nanoparticle electron microscopy, focusing on transmission electron microscopy (TEM), high-resolution transmission electron microscopy (HRTEM), scanning transmission electron microscopy (STEM), and in situ TEM. The discussion is organized around the principal challenges in nanoparticle characterization, including particle detection, segmentation, morphology quantification, atomic-resolution restoration, defect identification, two-dimensional-to-three-dimensional structural inference, and analysis of dynamic processes in situ. We review computational approaches from conventional ML and convolutional neural networks to transformer architectures, self-supervised learning, foundation models, multimodal AI, and physics-informed learning. We further discuss integrating microscopy data with simulations, metadata, and autonomous experimentation to relate nanoparticle structure, dynamics, synthesis conditions, and functional properties. The advantages, limitations, benchmarking, and data requirements of current methodologies are critically assessed. Finally, emerging opportunities for foundation models, AI-guided microscopy, closed-loop experimentation, and autonomous materials discovery are discussed. By integrating advances across computer vision, materials informatics, and electron microscopy, this review highlights the role of AI in next-generation nanoparticle characterization and accelerated materials discovery.

## 1. Introduction

Nanoparticles constitute one of the most intensively investigated classes of materials owing to the unique physical and chemical properties that emerge at the nanoscale [1]. Their performance is governed not only by composition but also by particle size, morphology, crystallographic structure, surface chemistry, defect distributions, and interparticle interactions, all of which strongly influence their functional behavior and technological performance [2–6]. These characteristics influence a wide range of applications, including heterogeneous catalysis [2, 9], energy conversion and storage [3, 10], plasmonics [4], magnetic devices [5, 11], biomedical technologies [6], sensing platforms [7, 12, 12], and environmental remediation [8]. Consequently, accurate characterization of nanoparticle systems is essential for understanding structure–property relationships and guiding the rational design of advanced nanomaterials [2, 3, 5].

Electron microscopy has become one of the most powerful tools for nanoparticle characterization because it provides direct access to structural and chemical information across multiple length scales [14, 15]. Conventional transmission electron microscopy (TEM) enables the determination of particle size distributions, morphology, agglomeration state, and dispersion characteristics [8, 35, 36]. High-resolution TEM (HRTEM) extends this capability to the atomic scale, allowing the visualization of lattice fringes, crystal defects, interfaces, strain fields, and facet structures [14, 18, 19]. Complementary diffraction-based techniques, including four-dimensional scanning transmission electron microscopy (4D-STEM), scanning precession electron diffraction (SPED), and electron ptychography, further provide crystallographic, orientation, strain, and phase information from multidimensional diffraction datasets [20]. More recently, advances in *in situ* TEM have enabled the direct observation of dynamic phenomena such as nanoparticle growth, coalescence, surface diffusion, phase transformations, and degradation processes under realistic operating conditions [21 - 27]. Together, these techniques generate an unprecedented amount of structural information that is critical for both fundamental investigations and technological applications.

Despite the remarkable capabilities of modern microscopy, the interpretation of nanoparticle images remains a challenging and often labor-intensive task. Conventional analysis workflows frequently rely on manual measurements, threshold-based image processing, or handcrafted feature extraction methods that may be time-consuming, subjective, and difficult to scale to increasingly large datasets. These limitations become particularly pronounced in high-throughput experiments, atomic-resolution imaging, and time-resolved *in situ* studies, where thousands of images or video frames may be generated during a single experiment [28, 29, 30]. The growing volume and complexity of microscopy data have therefore created a strong demand for automated and reproducible analysis methodologies capable of extracting meaningful information efficiently and consistently.

The rapid development of artificial intelligence (AI), machine learning (ML), and deep learning (DL) has provided new opportunities to address these challenges. Initially adopted for image classification and feature recognition tasks, AI methodologies have evolved into powerful tools capable of performing automated particle detection, semantic and instance segmentation, atomic resolution denoising, defect identification, image restoration, structural reconstruction, and dynamic process analysis [31, 32]. More recently, advances in self-supervised learning, transformer architectures, foundation models, and multimodal learning have expanded the scope of AI-assisted microscopy beyond image interpretation toward increasingly sophisticated forms of scientific inference [16, 33-36]. Rather than merely identifying features within images, modern AI systems are beginning to establish connections between structure, processing history, and functional behavior, thereby contributing directly to materials discovery and optimization.

The application of AI to microscopy has grown at an exceptional pace during the past decade. As illustrated in **Figure 1**, the number of publications combining nanoparticle characterization, electron microscopy, and AI has increased dramatically, reflecting both the growing availability of microscopy datasets and the maturation of ML methodologies. This rapid expansion has produced a diverse and highly interdisciplinary body of literature spanning computer vision, materials science, electron microscopy, data science, and materials informatics. While numerous studies have demonstrated the potential of AI-assisted microscopy [28, 31, 33, 34, 37], the field remains fragmented, with methodologies often developed for specific imaging modalities, materials systems, or analytical tasks.

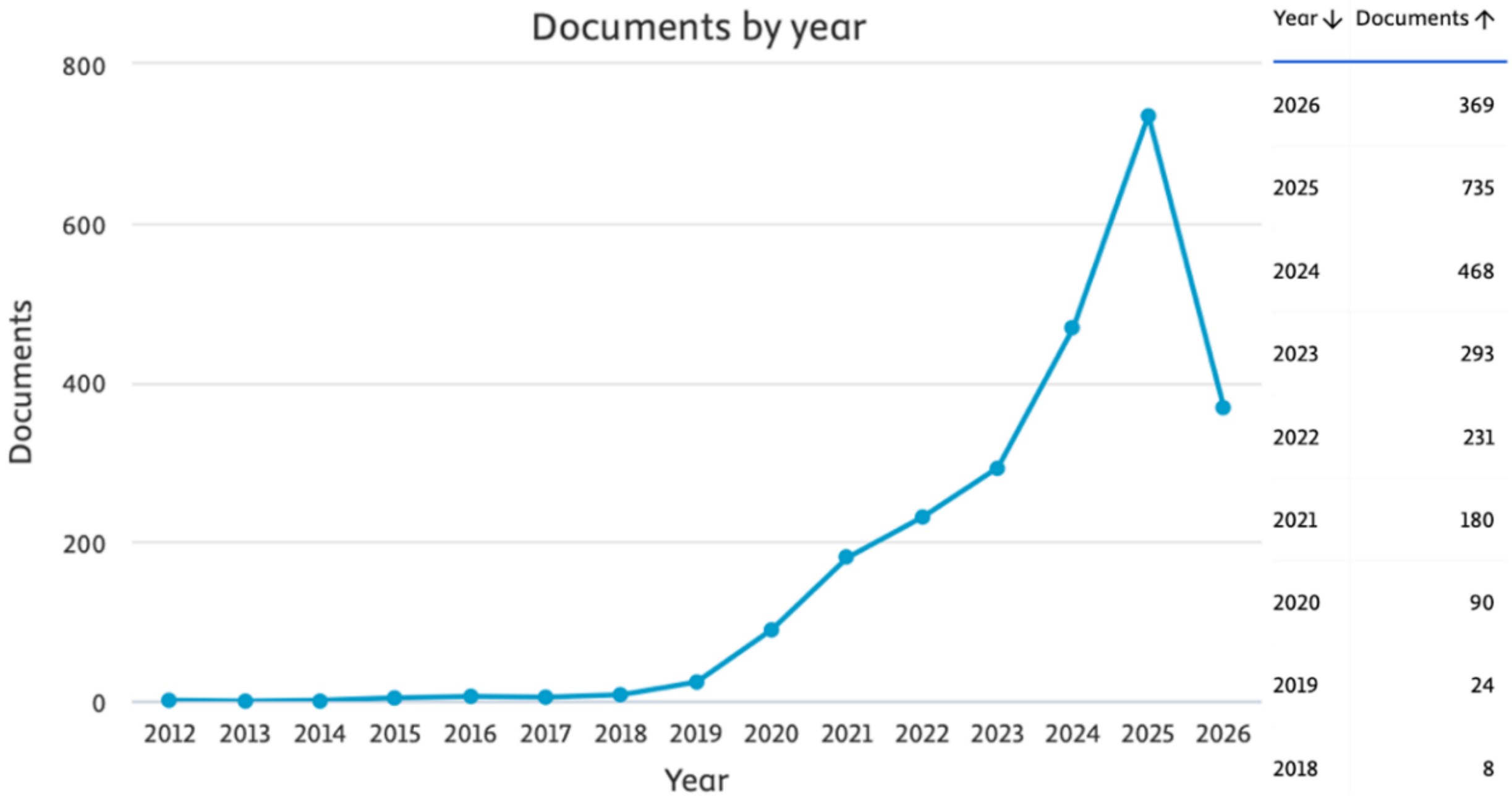

Figure 1. Annual number of publications related to artificial intelligence and nanoparticle electron microscopy indexed in Scopus between 2012 and 2026, based on the query *[(TEM OR HRTEM OR STEM) AND nanoparticle AND (deep learning OR machine learning)]* restricted to the Materials Science subject area. A total of 2415 documents were identified. The plot reveals limited activity during the early 2010s, followed by rapid expansion beginning around 2019 and accelerating strongly after 2021, with a peak output of 735 publications in 2025. This trend highlights the increasing adoption of machine learning and deep learning methods in nanoparticle microscopy, driven by advances in computer vision, growing microscopy datasets, and increased demand for automated quantitative image analysis. Data extracted from Scopus search analysis performed in April 2026.

This review provides a comprehensive overview of recent advances in artificial intelligence for nanoparticle microscopy, with particular emphasis on transmission electron microscopy, high-resolution TEM, and *in situ* TEM applications. The discussion is organized according to the principal scientific challenges encountered in nanoparticle characterization, including particle detection and segmentation, atomic resolution denoising and restoration, two-dimensional to three-dimensional structural inference, and spatiotemporal analysis of dynamic microscopy data. In addition to summarizing state-of-the-art methodologies, the review critically examines their performance, limitations, data requirements, and future prospects. Particular attention is devoted to emerging developments such as foundation models, self-supervised learning, multimodal AI systems, and autonomous microscopy workflows, which are expected to play an increasingly important role in the next generation of nanoparticle characterization platforms [35].

## 2. Electron Microscopy as a Source of Multiscale Information

Electron microscopy provides access to structural information spanning multiple spatial and temporal scales, making it one of the most powerful characterization techniques in nanoparticle science. Depending on the imaging mode, experimental conditions, and microscope configuration, electron microscopy datasets may contain information ranging from particle morphology and size distributions to atomic-scale structural features and time-resolved dynamic processes. The ability to directly visualize nanomaterials across these different levels of complexity has made electron microscopy indispensable for understanding nanoparticle formation, evolution, and functional behavior [14, 15, 38]. At the same time, the increasing quantity and complexity of microscopy data have created significant analytical challenges, motivating the development of AI methodologies capable of extracting meaningful information efficiently and reproducibly.

At the most fundamental level, electron microscopy provides morphological information regarding particle size, shape, spatial distribution, agglomeration behavior, and overall sample homogeneity. These descriptors are routinely used to evaluate synthesis quality, investigate growth mechanisms, and establish correlations between structural characteristics and material performance [2, 14]. Large nanoparticle ensembles are often analyzed statistically in order to determine size distributions, aspect ratios, particle densities, and dispersion characteristics, generating substantial quantities of image data [14]. Because such analyses frequently involve repetitive and time-consuming measurements, morphology-based characterization has become one of the earliest and most successful application areas of ML and DL, particularly for automated particle detection, counting, classification, and segmentation [28, 39].

Beyond morphology, high-resolution electron microscopy enables direct observation of atomic-scale structural features that often govern nanoparticle functionality. Lattice fringes, crystal defects, grain boundaries, interfaces, surface facets, dislocations, and local strain fields can all influence catalytic activity, magnetic properties, electronic transport, and mechanical behavior [14, 15, 40, 41]. Complementary diffraction-based techniques, including four-dimensional scanning transmission electron microscopy (4D-STEM), scanning precession electron diffraction (SPED), and electron ptychography, further extend electron microscopy by providing crystallographic, orientation, phase, and strain information from multidimensional diffraction datasets, creating additional opportunities for machine-learning-assisted analysis [15, 20]. The characterization of these features is therefore essential for understanding structure–property relationships in nanoscale materials. However, the extraction of atomic information is frequently complicated by imaging artifacts, limited signal-to-noise ratios, beam-induced damage, and contrast variations arising from microscope operating conditions. Consequently, substantial research efforts have focused on developing AI-assisted denoising, image restoration, defect identification, and structural interpretation methodologies capable of improving image quality while preserving physically meaningful information [28, 33].

A further level of complexity arises when attempting to infer three-dimensional structural information from inherently two-dimensional projections. Although electron microscopy images contain rich structural information, many nanoparticle characteristics of interest—including faceting, crystallographic orientation, internal defect distributions, and overall morphology—are fundamentally three-dimensional. Traditional approaches for obtaining such information often require tomography experiments or extensive image simulations, both of which can be resource-intensive and computationally demanding [43, 44]. Recent advances in DL, generative modeling, and physics-informed ML have demonstrated that meaningful structural inference can be extracted directly from two-dimensional microscopy data, enabling increasingly sophisticated reconstruction and interpretation of nanoparticle architectures [45, 46].

Modern electron microscopy has also evolved beyond static imaging through the development of *in situ* and *operando* methodologies that allow nanoscale processes to be observed under realistic experimental conditions. Through the use of specialized holders and environmental cells, researchers can directly monitor nanoparticle nucleation, growth, coalescence, diffusion, phase transformations, catalytic reactions, and degradation mechanisms while they occur [21, 22, 26, 27, 44]. These experiments generate large spatiotemporal datasets consisting of thousands or even millions of image frames, making manual analysis impractical. As a result, *in situ* microscopy has become a major driver for the development of AI methods capable of object tracking, event detection, trajectory analysis, temporal prediction, and autonomous experimental control [47].

The ultimate objective of AI-assisted microscopy extends beyond image interpretation toward scientific understanding and materials discovery. By integrating information extracted from images with complementary experimental measurements, simulations, and theoretical models, ML algorithms increasingly enable the identification of relationships between nanoparticle structure, dynamics, and functional properties [47, 48]. Such approaches provide opportunities to move beyond descriptive characterization and toward predictive frameworks capable of guiding synthesis optimization, accelerating materials development, and uncovering previously inaccessible structure–property relationships. This progression from image analysis to scientific inference represents one of the most significant ongoing transformations in electron microscopy.

**Figure 2** summarizes this hierarchy of information extraction in nanoparticle microscopy and the corresponding evolution of AI methodologies. The progression from morphology characterization and particle segmentation to atomic-scale interpretation, structural inference, dynamic process analysis, and mechanistic understanding reflects the increasing complexity of both the microscopy data and the scientific questions being addressed. The remainder of this review follows this framework, examining how ML and DL techniques have been developed to tackle each of these analytical challenges.

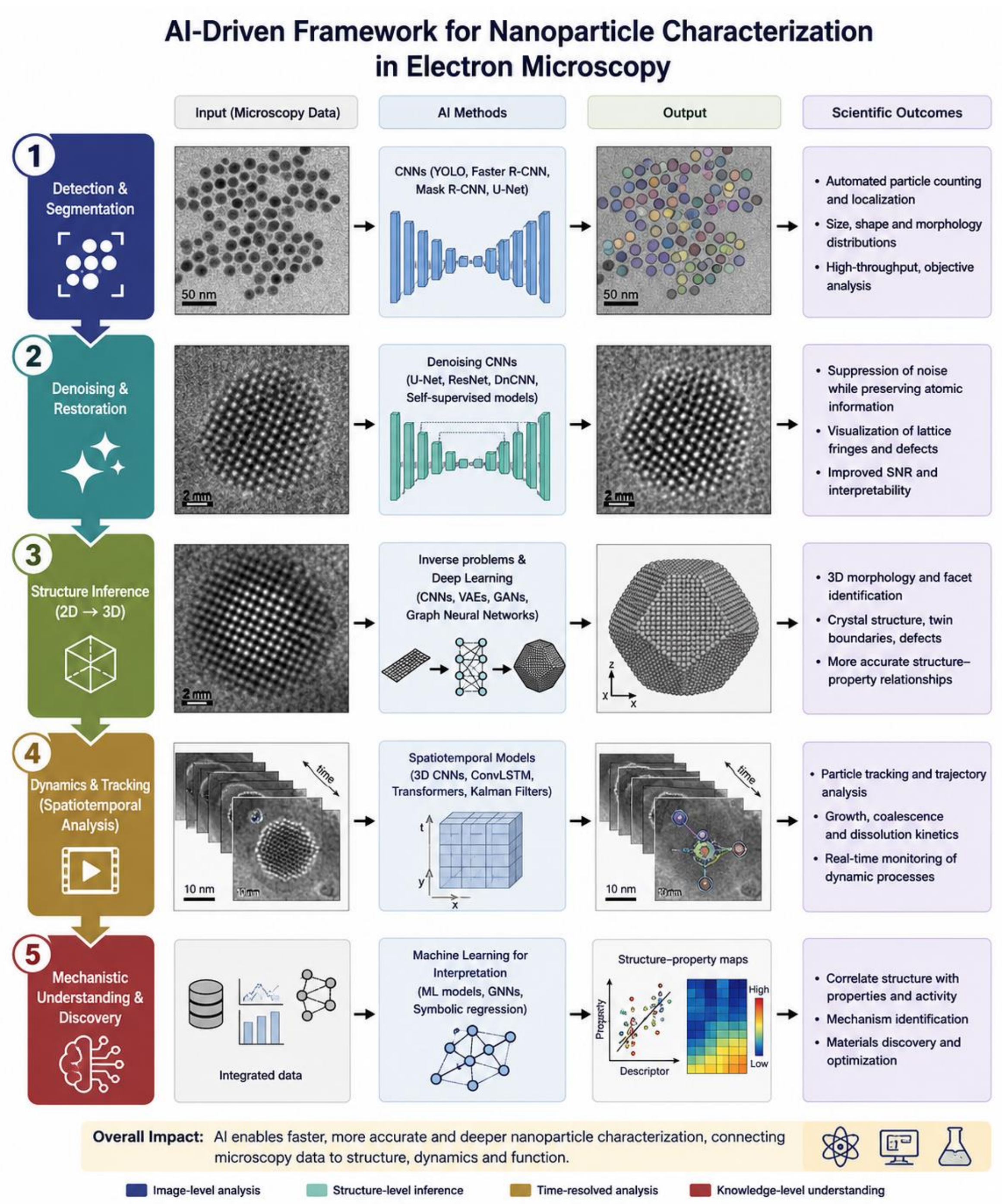

Figure 2. Hierarchical framework of DL tasks in nanoparticle microscopy, progressing from detection to spatiotemporal inference.

## 3. Deep Learning and Machine Learning Methods for Nanoparticle Microscopy

Importantly, the evolution of AI methodologies in nanoparticle microscopy has paralleled broader developments in computer vision and ML, with many of the architectures now routinely employed in microscopy originally developed for natural-image analysis and only subsequently adapted to scientific imaging problems. Early computational approaches relied on manually designed descriptors, including intensity statistics, shape parameters, texture features, Fourier transforms, and other handcrafted image representations [49]. These descriptors were subsequently analyzed using conventional ML algorithms such as support vector machines [50], decision trees, and random forests [51]. While effective for relatively simple and well-controlled datasets, such approaches often struggled to generalize across variations in contrast, noise levels, imaging conditions, sample thickness, and microscope settings that are commonly encountered in electron microscopy.

The type of information accessible to DL models depends strongly on the microscopy modality, as conventional TEM, HRTEM, and *in situ* TEM each provide distinct structural and temporal information at different length scales (**Figure 3**). Consequently, the evolution of AI methods in nanoparticle microscopy has closely followed the increasing complexity of the information being extracted. Early efforts focused primarily on automating repetitive image-analysis tasks, whereas more recent studies increasingly aim to infer structural, physical, and dynamic information directly from microscopy data [28].

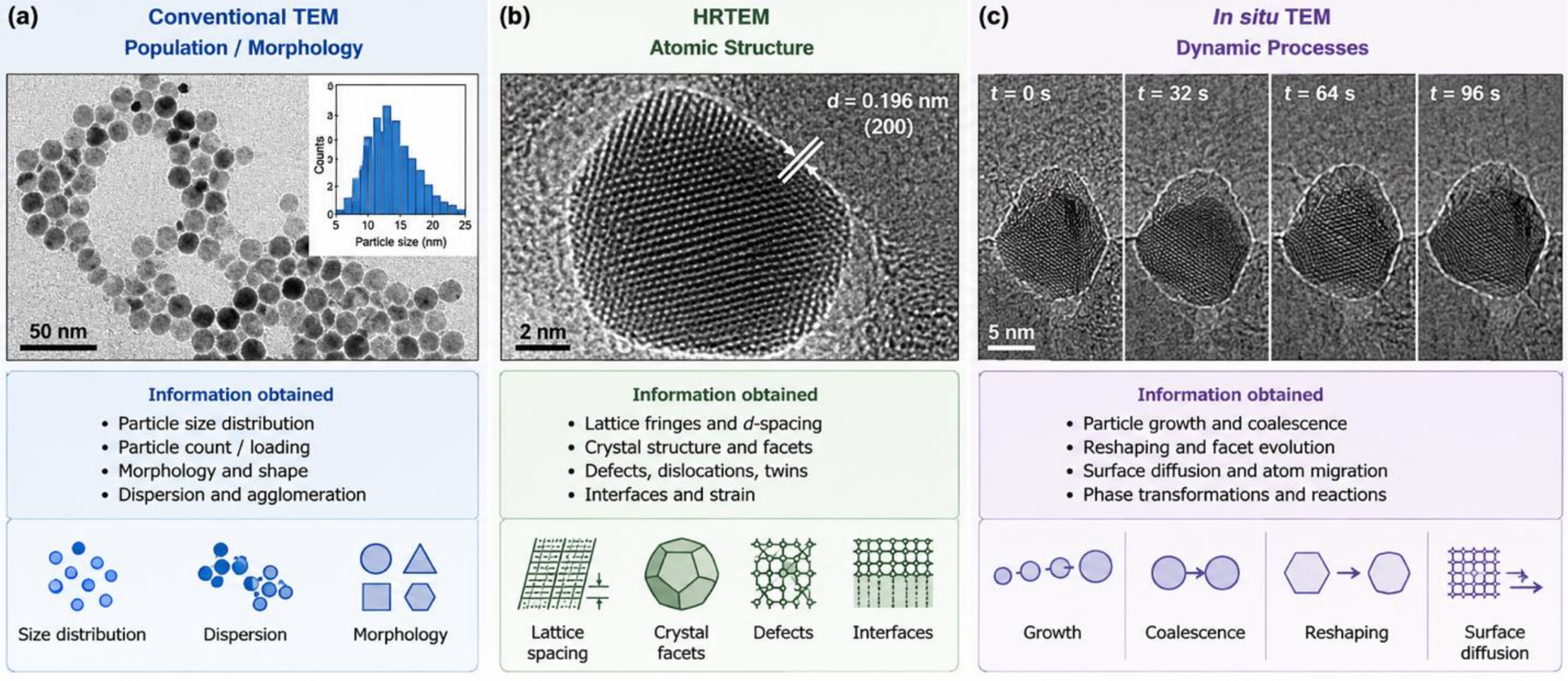

Figure 3. Representative electron microscopy modalities commonly used in nanoparticle characterization. (a) Conventional TEM illustrating nanoparticle ensemble morphology and size distribution. (b) HRTEM illustrating atomic-resolution lattice imaging within an individual nanoparticle. (c) Illustrative *in situ* TEM time sequence showing structural evolution during dynamic nanoparticle processes. All panels are author-generated schematic illustrations inspired by representative examples from the literature [14, 15, 21].

The introduction of convolutional neural networks (CNNs) marked a decisive turning point in microscopy image analysis by enabling hierarchical feature extraction directly from raw image data [52]**.** Unlike traditional workflows that depend on manually engineered descriptors, CNNs learn increasingly complex image representations during training, allowing them to identify features that may not be apparent through conventional analysis. This capability proved particularly valuable for nanoparticle characterization, where particle morphology, agglomeration state, crystallinity, and defect structures can exhibit significant variability. As a result, CNN-based approaches rapidly became the dominant paradigm for classification, detection, and segmentation tasks across a wide range of microscopy modalities [53, 54, 55]. Recent years have further expanded this landscape through vision transformers, self-supervised learning, diffusion models, and foundation models capable of generalizing across microscopy tasks [36, 56-67]. A major factor contributing to the success of deep learning in microscopy has been its ability to exploit large image datasets while maintaining robustness to experimental variability. Advances in transfer learning further accelerated adoption by allowing models pretrained on large image repositories to be adapted to specialized microscopy applications using comparatively limited datasets [68]. These developments enabled automated workflows capable of processing thousands of images with minimal user intervention, significantly improving throughput and reproducibility [47]. Consequently, DL has evolved from an experimental computational tool into a routine component of quantitative nanoparticle characterization.

As applications expanded from morphology analysis toward atomic-resolution imaging, DL methods became increasingly intertwined with the physics of image formation. In HRTEM and STEM, image contrast arises from complex interactions between the electron beam and the specimen and is strongly influenced by factors such as defocus, aberrations, thickness, and orientation. Under these conditions, DL models are frequently trained not only to recognize image features but also to recover physically meaningful structural information. This shift has led to growing interest in physics-informed learning strategies that incorporate prior knowledge of imaging mechanisms, crystallography, or simulation-based constraints into the training process [45, 47]. The scarcity of experimentally verified ground-truth data at atomic resolution has further encouraged the use of synthetic training datasets generated through multi-slice calculations and other forward models of electron scattering. By learning from large numbers of simulated images spanning different structural and imaging conditions, neural networks can establish relationships between observed contrast and underlying atomic configurations. Such simulation-driven methodologies have become increasingly important for tasks including lattice analysis, defect identification, structural reconstruction, and shape inference [45, 46]. More broadly, they represent a transition from purely image-based interpretation toward AI frameworks capable of approximating inverse problems in microscopy.

In parallel with advances in structural analysis, recent years have witnessed growing interest in the integration of deep learning with dynamic microscopy experiments. The increasing availability of *in situ* and *operando* electron microscopy has generated large spatiotemporal datasets that contain both structural and temporal information [22, 30]. Unlike conventional image-analysis problems, these datasets require models capable of simultaneously tracking objects, identifying events, quantifying temporal evolution, and predicting future behavior. This challenge has stimulated the development of spatiotemporal learning approaches that combine image-processing capabilities with temporal modeling, enabling the analysis of nanoparticle growth, coalescence, diffusion, phase transformations, and other dynamic processes [21, 22]. Another important indication of the maturation of the field is the growing translation of AI methodologies from academic research into commercial microscopy platforms. Automated particle detection, segmentation, sizing, classification, and workflow optimization are increasingly being incorporated into user-facing software packages, reducing the need for extensive manual intervention [28]. This trend reflects a broader shift in electron microscopy from proof-of-concept algorithm development toward integrated analytical ecosystems in which AI functions as a routine component of the characterization workflow.

Taken together, the literature reveals a clear evolution from image-level analysis toward structure-level interpretation, dynamic understanding, and ultimately scientific inference. Deep learning is no longer employed solely to identify objects within microscopy images but increasingly serves as a tool for extracting physically meaningful information, reconstructing hidden structural features, analyzing complex temporal behavior, and linking experimental observations to materials functionality. These developments provide the methodological foundation for the application-focused sections that follow, which examine detection and segmentation, denoising and restoration, structural inference, dynamic analysis, and mechanistic understanding in greater detail.

## 4. Detection and Segmentation in TEM Nanoparticle Imaging

Detection and segmentation constitute the most mature and widely adopted applications of AI in nanoparticle microscopy [28, 69]. These tasks operate primarily at the level of conventional TEM and low-resolution STEM, where the principal objective is to extract statistically meaningful descriptors of nanoparticle populations, including particle count, size distributions, morphology, and spatial organization. Unlike atomic-resolution analysis, where interpretation is strongly influenced by the physics of image formation, detection and segmentation rely primarily on geometric and intensity-based information, making them particularly well suited to data-driven learning approaches. Consequently, this area has served as an important proving ground for the application of ML and DL in materials characterization.

### 4.1 Nanoparticle Detection

Detection-based approaches aim to localize nanoparticles within microscopy images, typically through bounding-box representations that identify the position and approximate extent of individual objects. The adoption of object-detection architectures from computer vision, particularly the YOLO and Faster R-CNN families, has transformed nanoparticle characterization by enabling automated analysis of large image datasets [70, 71]. Recent studies have also demonstrated automated catalyst-particle analysis, polymer nanoparticle segmentation, boundary-aware networks, and high-throughput morphology pipelines [11, 16, 29, 72]. These methods are capable of rapidly identifying hundreds or thousands of nanoparticles within a single image while maintaining high levels of reproducibility and consistency. The growing popularity of detection-based workflows reflects the increasing need for high-throughput characterization in nanomaterials research. Automated detection enables rapid extraction of particle counts, size distributions, and spatial statistics that would otherwise require extensive manual effort. Recent studies have demonstrated the applicability of detection networks across a wide range of nanoparticle systems, including metallic nanoparticles, polymer nanoparticles, catalyst supports, and nanostructured thin films [39]. The success of these approaches highlights the ability of DL models to remain robust under variations in particle size, morphology, and imaging conditions while significantly reducing analysis time. Despite these advantages, detection methods inherently provide only coarse geometric information. Bounding-box representations are often sufficient for particle counting and localization but cannot accurately capture particle boundaries, surface features, or irregular morphologies. Consequently, detection is frequently employed as the first stage of more sophisticated image-analysis pipelines in which precise particle characterization is achieved through segmentation.

### 4.2 Nanoparticle Segmentation

Segmentation methods address the limitations of detection by assigning class labels at the pixel level, enabling accurate delineation of nanoparticle boundaries and more precise extraction of morphological descriptors. Unlike detection algorithms that localize objects using bounding boxes, segmentation networks directly identify the pixels belonging to individual nanoparticles, allowing quantitative measurements of particle area, perimeter, aspect ratio, circularity, surface curvature, and other shape-related characteristics [16]. Encoder–decoder architectures, particularly U-Net and its numerous derivatives, have become the dominant framework for nanoparticle segmentation [55]. Their success stems from their ability to simultaneously capture local image features and global contextual information through multi-scale feature extraction and skip connections. These characteristics are particularly important in microscopy images, where nanoparticles may exhibit substantial variability in size, contrast, orientation, and background environment. Segmentation networks have therefore become central components of automated TEM analysis workflows and are widely employed for quantitative nanoparticle characterization.

A representative workflow for AI-assisted nanoparticle analysis is illustrated in **Figure 4**. Raw microscopy images are first processed using detection or segmentation networks to identify individual particles and delineate their boundaries. The resulting masks are subsequently converted into quantitative descriptors, enabling automated extraction of particle statistics and morphology distributions. Such workflows represent a major advance over traditional manual analysis by providing improved scalability, reproducibility, and throughput.

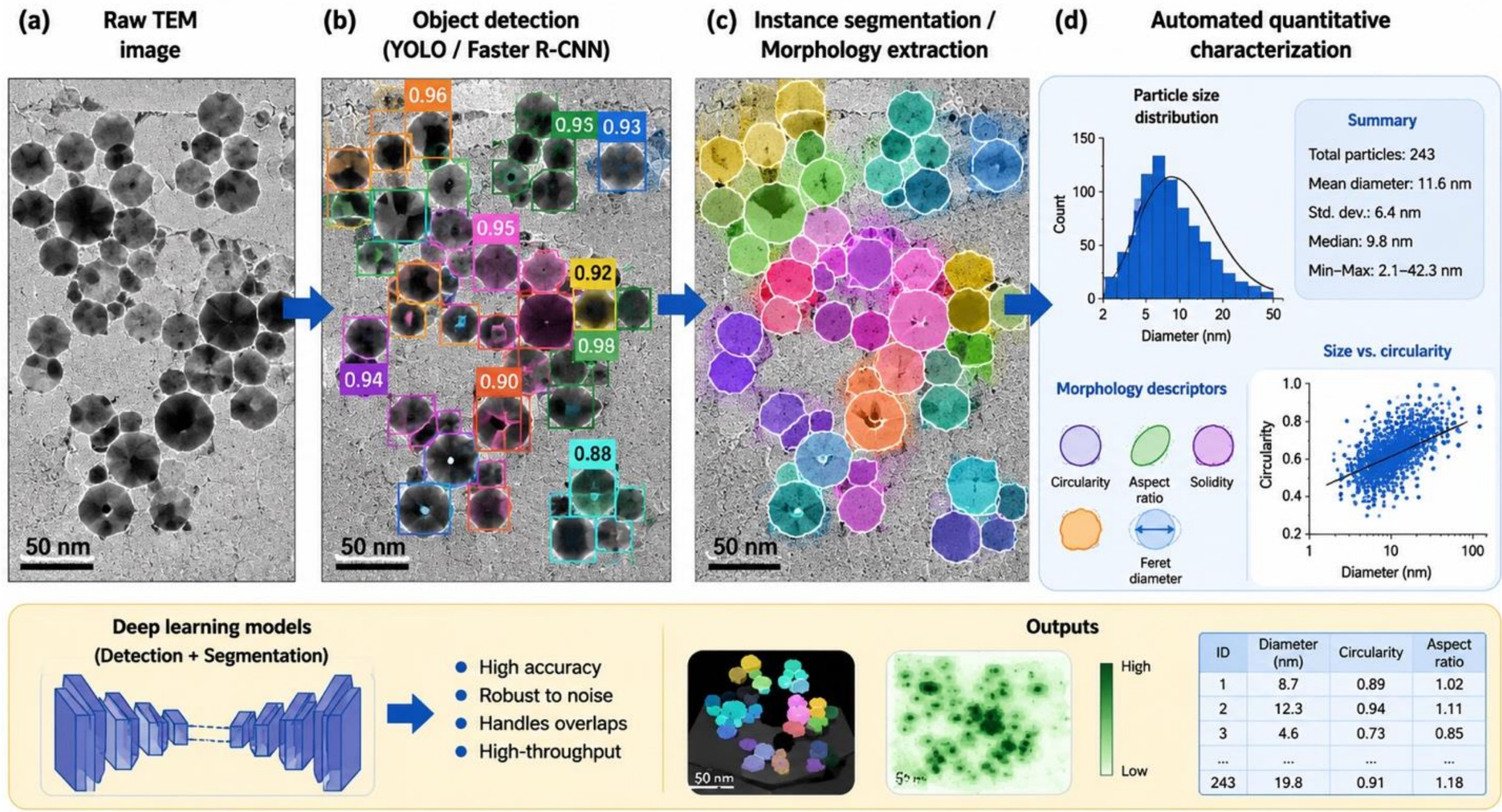

Figure 4. Workflow for AI-assisted nanoparticle analysis in TEM. Raw TEM images are processed through DL–based detection and instance segmentation to identify individual nanoparticles, delineate particle boundaries, and extract quantitative descriptors such as size distributions, morphology, and population statistics. The figure illustrates the transition from microscopy image acquisition to automated high-throughput nanoparticle characterization. Inspired by recent TEM analysis pipelines including DetectNano [39].

A major trend in recent years has been the transition from semantic segmentation toward instance segmentation [73]. While semantic segmentation identifies pixels belonging to nanoparticles as a single class, instance segmentation additionally separates individual particles into distinct objects. This distinction is particularly important for densely packed systems, catalyst nanoparticles, and agglomerated assemblies where particle overlap is common [16]. By simultaneously combining detection and segmentation, instance-segmentation approaches enable higher-fidelity characterization of complex nanoparticle populations and facilitate more accurate extraction of particle-specific morphological descriptors.

### 4.3 Foundation Models for Nanoparticle Segmentation

Recent developments in artificial intelligence have introduced foundation models capable of performing segmentation tasks across a wide range of image domains with limited task-specific training [74]. Among these, the Segment Anything Model (SAM) and its derivatives have attracted significant attention due to their ability to generate segmentation masks from simple prompts [75, 76]. Unlike conventional segmentation networks that require extensive retraining for each new dataset, foundation models offer improved flexibility and transferability across different imaging conditions and materials systems. The application of foundation models to nanoparticle microscopy remains an active area of research. Early studies suggest that SAM-based approaches can provide competitive segmentation performance while reducing annotation requirements and facilitating adaptation to previously unseen datasets [77, 78]. However, their performance on highly specialized nanoparticle images may still lag behind carefully trained task-specific architectures such as U-Net or nnU-Net [55, 79]. Consequently, a trade-off currently exists between the superior generalization capabilities of foundation models and the higher accuracy often achieved by specialized networks trained on curated microscopy datasets [61, 77, 78].

### 4.4 Challenges and Robustness

Despite their success, detection and segmentation methods remain sensitive to several challenges commonly encountered in electron microscopy. Variations in image contrast, particle overlap, heterogeneous supports, contamination, and imaging artefacts can all negatively affect performance. These challenges are often highly material-dependent. Soft-matter systems frequently exhibit weak contrast and diffuse boundaries, whereas metallic nanoparticles may appear as strongly overlapping high-contrast features [16, 39]. Similarly, catalyst nanoparticles supported on porous materials often present highly complex backgrounds that complicate automated analysis [16, 80]. To address these issues, recent studies have increasingly incorporated data augmentation, synthetic training datasets, domain adaptation strategies, and self-supervised learning techniques [35]. These approaches aim to improve generalization and reduce dependence on large quantities of manually annotated training data. Nevertheless, annotation remains one of the primary bottlenecks in nanoparticle segmentation, particularly for complex systems requiring expert interpretation.

### 4.5 Evaluation Metrics and Comparative Performance

Among all AI applications in nanoparticle microscopy, detection and segmentation represent the most mature and extensively benchmarked research areas. Their development has benefited from evaluation frameworks inherited from computer vision and biomedical image analysis, allowing quantitative comparison between different algorithms and datasets [55, 69, 70]. Nevertheless, performance assessment remains challenging because reported results often depend strongly on nanoparticle morphology, imaging conditions, dataset composition, and annotation quality.

For object detection, the most widely used metrics are precision, recall, and mean Average Precision (mAP). Precision quantifies the fraction of detected objects that correspond to real nanoparticles, whereas recall measures the fraction of nanoparticles successfully identified by the model [70]. Mean Average Precision combines these aspects across multiple confidence thresholds and has become the standard benchmark for detection algorithms [69]. For segmentation tasks, overlap-based metrics such as the Dice Similarity Coefficient (Dice) and Intersection-over-Union (IoU) are most commonly employed [55]. These metrics evaluate the agreement between predicted particle masks and reference annotations and directly influence the accuracy of downstream morphological measurements.

As shown in **Table 1**, modern detection and segmentation architectures routinely achieve high levels of performance on curated datasets. Dice coefficients above 0.90 and detection accuracies exceeding 90% are increasingly common, particularly when large, annotated datasets are available. However, caution is required when comparing reported values across studies. Performance may vary significantly depending on particle morphology, degree of agglomeration, image contrast, support materials, and annotation protocols. Furthermore, high numerical scores do not necessarily guarantee scientifically meaningful interpretation, as even small segmentation errors can influence measurements of particle curvature, facet distributions, and projected area.

Table 1. Representative performance metrics reported for nanoparticle detection and segmentation.

| Method Family | Representative Architecture | Task | Typical Metric | Typical Performance |
|---|---|---|---|---|
| Detection | YOLOv8 | Nanoparticle detection | mAP | >0.90 |
| Detection | Faster R-CNN | Nanoparticle detection | mAP | 0.85–0.95 |
| Segmentation | U-Net | Particle segmentation | Dice, IoU | 0.85–0.95 |
| Segmentation | nnU-Net | Particle segmentation | Dice, IoU | Frequently >0.90 |
| Foundation Models | SAM / MedSAM | Segmentation | Dice, IoU | ~0.80–0.90 |

The progression from detection to segmentation represents an important conceptual transition in AI-assisted microscopy. Detection addresses the question of where nanoparticles are located, whereas segmentation determines their precise boundaries and morphology. Together, these methodologies provide the quantitative descriptors required for higher-level analytical tasks. However, they remain fundamentally limited to features directly visible in two-dimensional images. This limitation motivates the developments discussed in the following sections, where DL is employed not only to identify observable structures but also to recover information obscured by noise, infer hidden structural features, and reconstruct three-dimensional characteristics from two-dimensional microscopy data.

## 5. Atomic-Resolution Denoising and Restoration in Nanoparticle Microscopy

Atomic-resolution denoising and restoration occupy a unique position within the broader landscape of AI for nanoparticle microscopy because they address a fundamentally different scientific problem from detection and segmentation. Whereas conventional TEM analysis is primarily concerned with identifying and quantifying visible particles, HRTEM and atomic-resolution STEM aim to recover physically meaningful structural information from images whose contrast is governed by electron scattering, phase interference, specimen thickness, aberrations, defocus, detector response, and electron-dose limitations. Consequently, the objective extends well beyond locating a particle or outlining its boundary: it involves preserving and recovering lattice fringes, atomic columns, surface reconstructions, grain boundaries, interfaces, defects, and local structural distortions that ultimately determine nanoparticle functionality. Unlike conventional image enhancement, atomic-resolution restoration cannot be regarded simply as a computational preprocessing step [67, 81, 82]. The restored image frequently serves as the basis for crystallographic interpretation, defect identification, strain analysis, and structure-property correlations. Any modification introduced by the restoration algorithm may therefore directly influence subsequent scientific conclusions. This distinction makes structural fidelity—not visual appearance—the central criterion for evaluating deep learning approaches in atomic-resolution microscopy. Recent developments have further extended AI-assisted interpretation to lattice-spacing extraction, atomic strain mapping, exit-wave reconstruction, and chemically resolved STEM analysis [18, 19, 67, 83, 84].

### 5.1 From Classical Filtering to Deep Learning Restoration

The rapid development of AI-assisted denoising has been driven largely by practical limitations inherent to electron microscopy itself. Many technologically important nanoparticle systems—including beam-sensitive catalysts, low-dimensional materials, soft matter, and ultrafine nanoclusters—cannot tolerate prolonged electron irradiation without structural modification [28, 22, 69, 85, 86, 87]. Under these conditions, low-dose imaging becomes essential, but the resulting images frequently suffer from poor signal-to-noise ratios that obscure atomic-scale information. Traditional denoising approaches, including Fourier masking, Gaussian filtering, Wiener filtering, and non-local means algorithms, have long been employed to improve image quality [87, 88]. While often effective in reducing visible noise, these methods may simultaneously suppress genuine structural information or introduce artifacts that complicate scientific interpretation. Deep learning approaches have transformed this landscape by learning complex nonlinear mappings between noisy and high-quality images, enabling substantially improved recovery of lattice-level information while preserving fine structural detail [31, 69, 89].

Importantly, the central scientific question rapidly evolved from whether neural networks could produce visually cleaner images to whether they could recover physically meaningful structural information without introducing misleading artifacts. This conceptual transition has become one of the defining themes of AI-assisted atomic-resolution microscopy and underlies much of the recent literature in the field. The typical impact of deep learning restoration on nanoparticle HRTEM analysis is illustrated schematically in **Figure 5**.

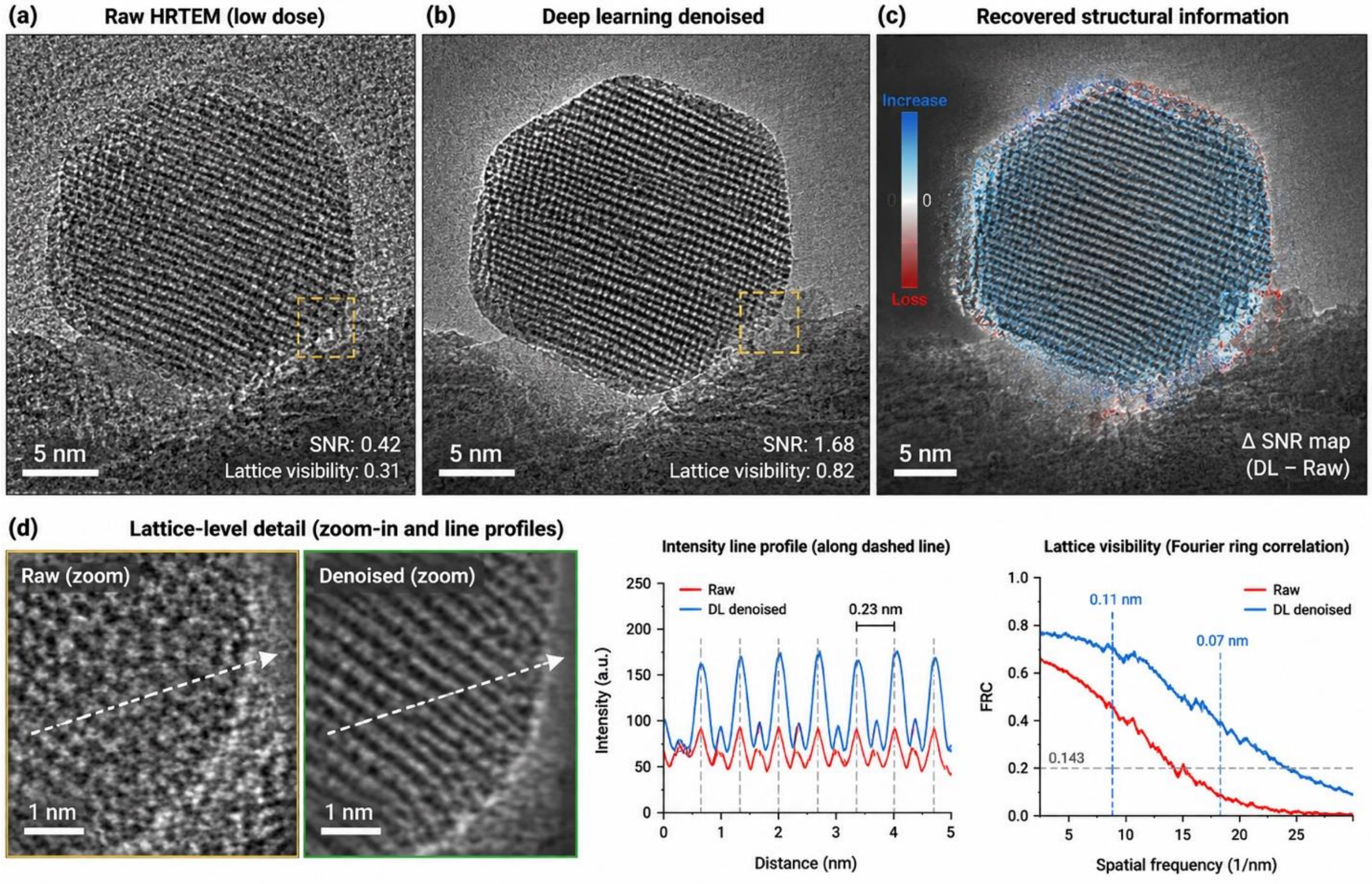

Figure 5. Deep learning-assisted restoration of atomic-resolution nanoparticle images. (a) Representative low-dose HRTEM image with limited signal-to-noise ratio. (b) Deep-learning-based restoration revealing enhanced lattice visibility while preserving structural information. (c) Difference map illustrating recovered structural information concentrated around lattice planes, particle surfaces, and interfaces. (d) Enlarged comparison together with representative intensity profiles and Fourier-based lattice visibility analysis, demonstrating the importance of structural fidelity rather than purely visual enhancement. All panels are author-generated schematic illustrations inspired by recent advances in AI-assisted denoising for electron microscopy [31, 69, 89].

## 5.2 Evolution Toward Physically Meaningful Restoration

One of the clearest trends emerging from recent literature is the gradual transition from generic image denoising toward physically informed restoration. Early studies demonstrated that CNNs could successfully recover lattice information and assist atomic-scale measurements, establishing the feasibility of learned restoration for HRTEM analysis [18, 19, 89]. Subsequent work increasingly emphasized interpretability and structural validation, recognizing that numerical image quality alone provides insufficient evidence of scientific reliability [28, 33, 69]. A particularly important development has been the growing integration of microscopy physics into deep learning workflows. Rather than treating restoration as a purely visual enhancement problem, modern approaches increasingly attempt to recover quantities more directly related to electron scattering or image formation itself, including exit-wave reconstruction and physically constrained representations [20, 33, 35]. This shift reflects a broader evolution in which AI is used not simply to improve image appearance but to recover information that more faithfully represents the underlying specimen.

Closely related to this trend is the increasing use of simulation-driven supervision. Because experimentally acquired noise-free atomic-resolution images rarely exist, many restoration models are trained using multi-slice simulations or other forward models capable of generating physically realistic microscopy data under controlled imaging conditions [37, 45]. Such approaches allow systematic variation of particle orientation, specimen thickness, substrate effects, aberrations, detector characteristics, and electron dose, creating extensive training datasets that would be impossible to obtain experimentally. This methodology provides a significant advantage unique to electron microscopy: unlike conventional computer vision applications, training data can be generated directly from physically modeled atomic structures. At the same time, however, restoration quality becomes critically dependent on the realism of the underlying simulations. Inadequate representation of substrate disorder, experimental variability, detector response, or structural complexity may produce models that perform well under validation conditions yet generalize poorly to real experimental images.

### 5.3 Denoising as an Enabling Technology for Scientific Discovery

Another major trend evident throughout the literature is the transformation of denoising from an isolated image-processing task into an enabling technology for downstream scientific inference. Restoration is increasingly integrated with defect identification, crystallographic analysis, phase recognition, local chemistry mapping, structural reconstruction, and three-dimensional inference, forming part of larger AI-assisted analytical pipelines rather than serving as a standalone preprocessing step [18, 83, 13]. Perhaps the most striking demonstrations of this transition have emerged from recent work on *in situ* electron microscopy [20, 22]. Deep learning-based video denoising frameworks have enabled recovery of atomic-resolution information directly from extremely noisy experimental recordings acquired under low-dose conditions, making previously inaccessible nanoscale dynamics experimentally observable. Rather than merely producing visually cleaner movies, these methods have revealed transient surface restructuring, adlayer formation, atomic migration, local strain propagation, and nanoparticle fluxionality occurring on millisecond timescales [20, 22]. The scientific significance of these developments extends well beyond image quality improvement. AI-assisted restoration increasingly enables direct observation of physical phenomena that would otherwise remain hidden beneath experimental noise, thereby expanding the observable space of electron microscopy itself. In this sense, denoising is evolving from an image-enhancement technique into a genuine tool for scientific discovery. In parallel, improvements in detector technology, sparse acquisition strategies, and Bayesian compressive sensing have increasingly complemented DL-based restoration methods [84 - 86].

### 5.4 Evaluation Metrics, Structural Fidelity, and Benchmarking

The evaluation of atomic-resolution denoising differs fundamentally from that of detection and segmentation because, in most practical microscopy experiments, an experimentally verified noise-free ground truth does not exist. Beam damage, dose limitations, detector noise, specimen drift, and temporal constraints frequently prevent acquisition of ideal reference images, making direct comparison between prediction and reality impossible [28, 69]. Historically, restoration quality has often been evaluated using image-processing metrics such as Peak Signal-to-Noise Ratio (PSNR) and Structural Similarity Index Measure (SSIM) [90]. When simulated reference data are available, these quantities provide useful numerical measures of reconstruction quality. However, their relevance to experimental microscopy remains inherently limited because visually similar images do not necessarily preserve physically meaningful structural information.

For atomic-resolution nanoparticle characterization, scientifically important information is frequently encoded within subtle lattice distortions, point defects, twin boundaries, surface reconstructions, stacking faults, or local disorder. Restoration algorithms may therefore achieve excellent PSNR or SSIM values while simultaneously introducing artificial crystallographic features or suppressing genuine structural information [69, 89]. As a consequence, increasing emphasis has been placed on microscopy-specific validation methodologies based on structural fidelity rather than image appearance alone. Among these approaches, Fourier Ring Correlation (FRC) has emerged as one of the most informative quantitative metrics because it evaluates preservation of spatial-frequency information and provides an estimate of effective image resolution across multiple length scales [91]. Unlike purely pixel-based measures, FRC correlates more directly with the preservation of crystallographic information and has therefore become increasingly important for evaluating AI-assisted restoration in electron microscopy. Representative evaluation methodologies associated with various methods for atomic-resolution denoising and restoration are tabulated in **Table 2**.

Table 2. Representative evaluation methodologies for atomic-resolution denoising and restoration.

| Method Family | Representative Approach | Typical Metrics | Primary Validation |
|---|---|---|---|
| CNN-based restoration | U-Net variants | PSNR, SSIM | Simulated ground truth |
| Physics-informed restoration | Exit-wave reconstruction | FRC, spatial resolution | Structural consistency |
| Noise-aware segmentation | Deep segmentation under varying noise | Dice robustness | Noise sensitivity analysis |
| Self-supervised restoration | Blind-spot architectures | SSIM, FRC | Experimental validation |
| Deep video denoising | Unsupervised video denoising | SNR improvement, FRC, temporal consistency | Dynamic experimental observations |

The evolution of benchmarking strategies reflects a **broader conceptual shift** within the field. Earlier work primarily emphasized numerical image-quality measures, whereas contemporary studies increasingly evaluate whether restored images preserve physically meaningful crystallographic information and support reliable downstream scientific interpretation [37, 35]. This transition from visual enhancement toward structural validation represents one of the defining characteristics of modern AI-assisted atomic-resolution microscopy.

## 6. 2D-to-3D Structural Inference and Reconstruction in Nanoparticle Microscopy

Among all current applications of AI in nanoparticle microscopy, three-dimensional structural inference represents perhaps the most ambitious objective [33, 48]. Unlike detection, segmentation, or denoising—which operate on information already present in the recorded image—structural inference attempts to recover properties that are only indirectly encoded in the experimental projection. The task is therefore fundamentally an inverse problem, requiring the estimation of latent structural variables from incomplete observations [35, 37, 48]. Rather than simply improving image interpretation, these methods seek to reconstruct aspects of nanoparticle geometry, crystallography, and internal organization that cannot be directly measured from a single two-dimensional image.

A defining limitation of TEM is that it provides only a two-dimensional projection of inherently three-dimensional nanostructures. Although this projection contains rich structural information, recovering the full three-dimensional morphology of a nanoparticle from one or a small number of images remains fundamentally underdetermined. Classical approaches such as electron tomography address this problem by reconstructing volumetric information from tilt series acquired at multiple viewing angles, but these techniques require extensive data acquisition, relatively high electron doses, and often become impractical for beam-sensitive materials or dynamic experiments [25, 43, 44, 66]. Complementary approaches based on electron ptychography have emerged as powerful alternatives, enabling quantitative phase retrieval and high-resolution three-dimensional imaging while operating under comparatively low electron-dose conditions. By exploiting diffraction information collected during STEM, electron ptychography can recover structural information that is inaccessible using conventional imaging alone, making it particularly attractive for beam-sensitive nanomaterials and limited-dose experiments [92]. More recently, AI-driven adaptive acquisition strategies, including deep reinforcement learning for ptychographic scanning, have demonstrated the potential to optimize data collection in real time, reducing electron dose while maintaining reconstruction quality [72].

These developments further illustrate the growing convergence between advanced imaging physics and AI, where computational methods increasingly contribute not only to image interpretation but also to the acquisition process itself. These experimental limitations have stimulated growing interest in AI-driven methodologies capable of inferring three-dimensional information directly from limited two-dimensional observations [37, 45, 46]. Unlike image restoration, where successful recovery can often be evaluated through comparison with simulated or experimentally acquired references, three-dimensional structural inference is intrinsically ambiguous. Multiple physically distinct nanoparticle configurations may produce highly similar experimental projections, particularly under different orientations, thicknesses, and imaging conditions [37, 44, 45]. Consequently, no DL architecture can uniquely recover arbitrary three-dimensional structure from a single image without incorporating prior knowledge through simulations, physical constraints, or statistical assumptions encoded during training. Understanding these assumptions is therefore essential when interpreting the predictions generated by AI-based reconstruction methods.

### 6.1 Simulation-Driven Inverse Learning

The dominant paradigm for AI-assisted structural inference is based on simulation-driven learning. In this framework, physically realistic microscopy images are generated from known atomic structures using multi-slice calculations or related forward models, producing large datasets in which both the image and its underlying three-dimensional structure are known. Neural networks are then trained to learn the inverse mapping between image contrast and structural configuration, effectively approximating the inverse imaging problem through statistical learning. The general concept underlying AI-assisted 2D-to-3D structural inference is illustrated in **Figure 6**. Rather than detecting observable image features, deep learning models extract latent structural representations from HRTEM or STEM contrast and associate them with candidate three-dimensional morphologies, crystallographic orientations, facet distributions, or internal structural configurations. In this sense, the network does not simply recognize objects but attempts to infer hidden structural variables that are only indirectly encoded within the experimental image.

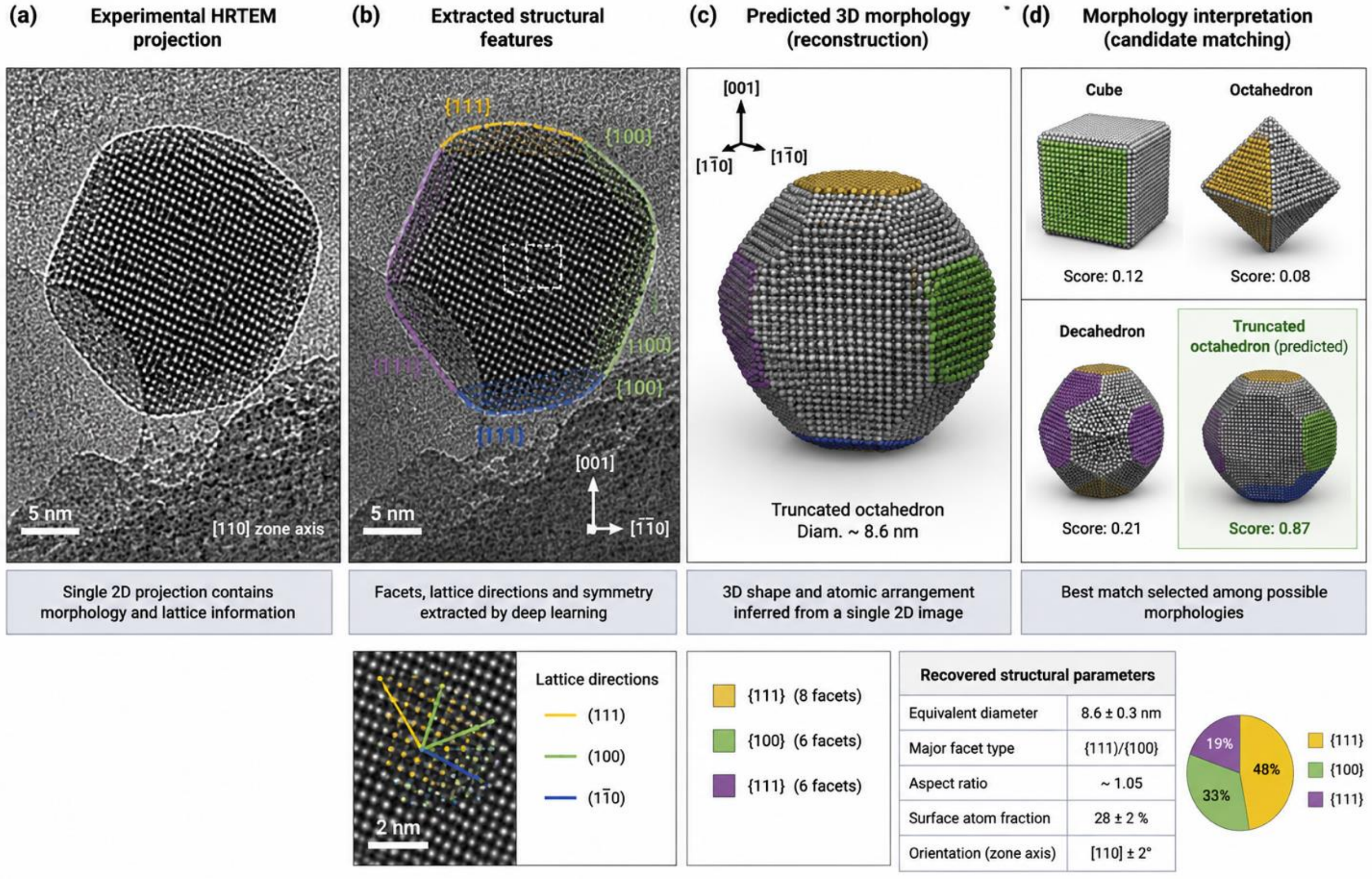

**Figure 6.** Schematic illustration of DL-based 2D-to-3D structural inference in nanoparticle HRTEM. (a) Representative 2D HRTEM projection containing lattice-level and morphological information. (b) Extraction of structural descriptors through DL analysis of local image contrast and atomic-scale patterns. (c) Reconstruction or classification of the corresponding three-dimensional nanoparticle morphology. (d) Recovery of physically meaningful descriptors including particle geometry, crystallographic orientation, exposed facets, and size-related parameters. All panels are author-generated schematic illustrations inspired by recent developments in AI-assisted nanoparticle structural inference [37, 45, 46].

Recent studies have demonstrated the feasibility of this strategy across a variety of applications, including sparse-view electron tomography, morphology classification from single HRTEM projections, atomic defect recognition, and chemically resolved STEM analysis [37, 45, 46]**.** Simulation-driven training has become particularly attractive because experimentally verified three-dimensional ground truth is rarely available, whereas synthetic datasets allow systematic variation of particle size, morphology, crystallographic orientation, substrate effects, imaging conditions, and electron dose under fully controlled conditions [37, 45]. These developments illustrate a broader transition in which DL is increasingly used to solve inverse physical problems rather than purely image-processing tasks [35, 48]. An important consequence of this methodology is that the quality of the inferred structure depends directly on the realism of the underlying simulations. If training datasets fail to adequately represent substrate disorder, detector response, aberration variability, beam effects, or structural heterogeneity, apparently impressive numerical performance may not translate into reliable experimental predictions. Consequently, simulation fidelity has become one of the principal determinants of model generalization and remains an active area of research [35, 37, 45]. Recent approaches combining multimodal learning with microscope metadata further suggest that structural inference may increasingly exploit experimental conditions in addition to image information alone [16, 35].

### 6.2 Beyond Classification: Continuous and Generative Structural Representations

Many early DL approaches formulated structural inference as a classification problem in which nanoparticles were assigned to predefined categories such as cubes, octahedra, decahedra, icosahedra, or truncated octahedra [45]. While highly effective for well-defined morphology libraries, such approaches inherently restrict predictions to discrete classes and may not adequately represent the continuous structural variability commonly encountered in real nanoparticle systems. Recent developments increasingly move beyond discrete classification toward continuous latent-space representations capable of encoding gradual variations in geometry, size, faceting, and atomic arrangement [35, 37]. Variational autoencoders, generative models, implicit neural representations, and related architectures provide flexible low-dimensional descriptions of structural space, allowing interpolation between known configurations and enabling inference beyond predefined categories. Although still emerging within nanoparticle microscopy, these methodologies represent a promising direction toward more realistic descriptions of structurally heterogeneous materials [42, 93, 94]. This transition is particularly important because nanoparticles frequently exhibit non-ideal morphologies, surface reconstructions, twin boundaries, strain-induced distortions, and size-dependent structural transitions that cannot easily be represented through simple categorical labels. Generative approaches therefore offer the possibility of representing structural uncertainty and continuous morphological evolution in a manner more closely aligned with the underlying physics of nanoscale systems [42, 93 - 96].

### 6.3 Evaluation, Benchmarking, and Uncertainty Quantification

If validating denoised images is challenging because experimentally noise-free references rarely exist, benchmarking three-dimensional structural inference presents an even greater difficulty. In this case, the objective is not merely to improve image quality but to recover information that is fundamentally absent from the original projection. Consequently, evaluation strategies differ substantially from those employed for detection, segmentation, or denoising and must explicitly account for the ambiguity inherent to inverse reconstruction problems. Many current studies employ classification-based evaluation frameworks in which predicted morphologies are compared against known structural labels using conventional metrics such as accuracy, precision, recall, F1-score, and confusion matrices [45]. These measures provide useful indicators of performance when models are trained and evaluated on consistent datasets and remain valuable for comparing alternative neural-network architectures under controlled conditions. However, classification accuracy alone provides only a partial description of structural inference quality, since correctly identifying an overall morphology does not necessarily imply accurate recovery of truncation degree, surface faceting, internal defects, or local atomic arrangement [37, 45].

For reconstruction-based approaches, benchmarking becomes even more complex. When models predict volumetric structures or atomic configurations, evaluation often relies on reconstruction error metrics such as root-mean-square error (RMSE), voxel similarity measures, or structural overlap indices relative to known simulated references [46] (see **Table 3**). While useful for methodological development, these metrics may substantially overestimate practical performance because corresponding experimental ground truth is rarely available under realistic microscopy conditions.

Table 3. Representative evaluation methodologies for AI-assisted three-dimensional structural inference.

| **Method Family** | **Representative Task** | **Typical Metrics** | **Principal Limitation** |
|---|---|---|---|
| Morphology classification | CNN-based classification | Accuracy, Precision, Recall, F1-score | Limited structural detail |
| Shape inference from HRTEM | Simulation-trained deep learning | Accuracy, Confusion Matrix | Domain shift |
| Sparse-view tomography | Deep learning reconstruction | RMSE, Reconstruction Error | Limited-angle dependence |
| Atomic configuration inference | Physics-informed deep learning | Structural overlap, Classification Accuracy | Experimental validation |
| Generative structural inference | Latent-space and generative models | Reconstruction Error, Similarity Metrics | Uncertainty quantification |

An additional challenge concerns the widespread reliance on synthetic datasets for both training and validation. Although simulation-based approaches provide access to large, labeled datasets and controlled structural variation, differences between simulated and experimental imaging conditions may significantly affect model generalization [35, 37, 45]. Variations in detector response, sample preparation, contamination, substrate disorder, and experimental noise frequently produce domain shifts that are not fully represented within simulated datasets. Consequently, benchmark values reported under idealized conditions should often be regarded as upper performance limits rather than realistic estimates of routine experimental performance. Growing attention is therefore being devoted to uncertainty estimation and probabilistic inference frameworks that explicitly acknowledge the ambiguity of the inverse problem [95, 96]. Rather than producing a single deterministic prediction, these approaches generate confidence estimates or probability distributions over candidate structures, providing users with more realistic assessments of prediction reliability. Hybrid simulation-experiment training strategies, physics-informed neural networks, Bayesian inference, and uncertainty-aware deep learning are likely to become increasingly important as AI-assisted structural inference moves toward practical scientific applications [35, 48, 95, 96].

### 6.4 From Image Interpretation to Scientific Inference

The progression from restoration to structural inference represents a fundamental conceptual transition in the role of AI within electron microscopy. Whereas denoising seeks to recover experimentally obscured information already present in the recorded signal, structural inference attempts to estimate latent physical variables that are only indirectly encoded in the image formation process. In this sense, DL increasingly functions not merely as an image-processing tool but as an inference engine capable of integrating experimental observations with prior physical knowledge [35, 48]. This evolution brings nanoparticle microscopy closer to the broader paradigm of scientific ML, in which experimental measurements, physical modeling, simulation, and statistical inference are combined within unified computational frameworks. In many recent studies, denoising, feature extraction, morphology classification, and structural inference already operate within integrated end-to-end pipelines that transform raw microscopy images into physically meaningful descriptors directly linked to catalytic performance, electronic behavior, mechanical response, or materials functionality [48]. At the same time, these developments emphasize the importance of rigorous validation, explainability, and uncertainty quantification. Because inferred structures may directly influence scientific interpretation and materials design decisions, confidence estimates and physics-based consistency checks are likely to become essential components of future AI-assisted reconstruction frameworks. Extending these concepts from static images to evolving image sequences naturally leads to the next frontier of AI-assisted nanoparticle microscopy: the analysis of dynamic nanoscale processes through spatiotemporal learning methods.

## 7. Spatiotemporal Deep Learning for *In Situ* TEM and Nanoparticle Dynamics

The extension of DL from static image analysis to spatiotemporal modeling represents one of the most significant recent developments in AI-assisted nanoparticle microscopy. Whereas conventional TEM and HRTEM primarily provide snapshots of nanoparticle structure, *in situ* and *operando* electron microscopy enable direct observation of processes evolving under external stimuli such as heating, gas exposure, liquid environments, electrical bias, or mechanical loading [21, 22]. Consequently, the objective shifts from characterizing static structures to understanding dynamic phenomena including nucleation, growth, coalescence, diffusion, phase transformations, surface restructuring, and degradation mechanisms. This transition fundamentally changes both the nature of the data and the role of AI. Instead of processing isolated images independently, DL models must analyze temporally correlated sequences, identify evolving structural features, and extract quantitative information describing the kinetics and mechanisms governing nanoparticle evolution. The resulting datasets frequently comprise thousands or even millions of image frames, making manual analysis impractical and motivating the development of increasingly sophisticated spatiotemporal learning frameworks [22, 28, 97].

### 7.1 From Frame-by-Frame Analysis to Spatiotemporal Learning

Early applications of DL to *in situ* microscopy largely extended detection and segmentation methodologies originally developed for static images [16, 17, 39, 70, 71, 73, 75, 76, 77, 79]**.** Convolutional neural networks successfully enabled automated identification of nanoparticles, defects, and interfaces across successive video frames, allowing trajectories and structural evolution to be reconstructed with greatly improved efficiency compared with manual analysis [16, 39]. Although highly successful, these approaches generally relied on independent frame analysis followed by *post hoc* tracking algorithms that associated objects between consecutive images. Self-supervised learning has recently enabled the analysis of liquid-phase TEM sequences without exhaustive manual annotation [65]. Such methodologies effectively quantified motion but only partially exploited the rich temporal information inherently present within microscopy videos.

More recent developments increasingly recognize that nanoparticle dynamics cannot be understood through independent frame analysis alone. Instead, temporal continuity itself contains physically meaningful information regarding growth mechanisms, particle interactions, diffusion pathways, and structural transformations. This realization has stimulated growing interest in architectures that explicitly model temporal dependencies, including recurrent neural networks, ConvLSTM frameworks, transformer-based sequence models, and hybrid spatial-temporal architectures capable of simultaneously learning spatial structure and temporal evolution [97, 98]. **Figure 7** illustrates the general concept of AI-assisted spatiotemporal analysis in *in situ* nanoparticle microscopy.

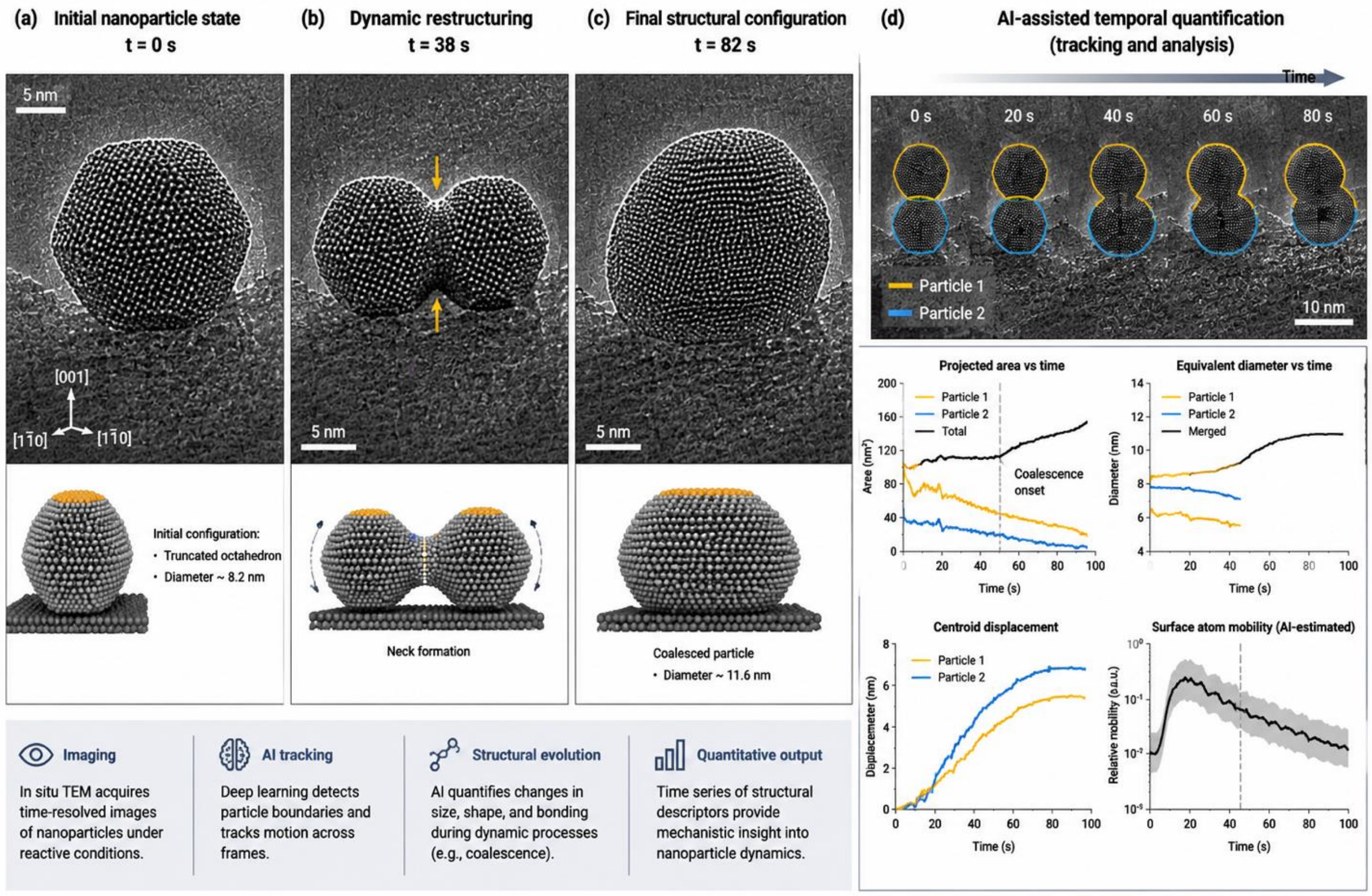

**Figure 7.** Schematic illustration of AI-assisted spatiotemporal analysis of nanoparticle dynamics in *in situ* TEM. (a) Initial nanoparticle configuration before structural evolution. (b) Dynamic restructuring and neck formation during particle interaction. (c) Final coalesced nanoparticle morphology. (d) Deep learning-assisted temporal quantification through automated tracking and extraction of projected area, equivalent diameter, centroid displacement, particle trajectories, and estimated surface mobility. The figure illustrates the transformation of time-resolved microscopy videos into quantitative descriptions of nanoscale dynamics and mechanistic evolution. All panels are author-generated schematic illustrations inspired by recent developments in AI-assisted *in situ* electron microscopy [22, 97, 98].

## 7.2 Deep Learning as an Enabler of Dynamic Observation

Perhaps the most profound recent development is that DL increasingly expands what can be experimentally observed rather than merely automating downstream image analysis. This distinction is particularly evident in low-dose and high-speed *in situ* microscopy, where experimental limitations frequently obscure transient structural states beneath severe noise [22, 93]. Recent advances in unsupervised deep video denoising have demonstrated that neural networks trained directly on noisy experimental sequences can recover atomic-resolution information without requiring clean reference data [31, 89]. Such approaches have enabled visualization of rapid surface restructuring, transient disorder, adlayer formation, surface atom migration, and nanoparticle fluxionality occurring on millisecond timescales under reactive environments [22]. Rather than simply improving image appearance, these methodologies effectively enlarge the experimental observability window itself, revealing nanoscale processes that would otherwise remain inaccessible. More broadly, these developments illustrate a conceptual transition in which AI becomes integrated directly into the experimental measurement process rather than functioning solely as a post-processing tool. In this sense, AI increasingly contributes not only to data interpretation but also to the practical limits of experimental observation.

### 7.3 Quantitative Analysis of Dynamic Processes

Beyond visualization, spatiotemporal DL enables extraction of quantitative descriptors describing nanoparticle evolution over time. Automated tracking of particle boundaries, centroid positions, projected area, equivalent diameter, local morphology, and interaction events provides access to detailed kinetic information that would be prohibitively time-consuming to obtain manually [16, 39]. Importantly, many scientifically relevant quantities emerge not from individual observations but from their temporal evolution. Growth rates, coalescence kinetics, diffusion coefficients, particle mobility, surface reconstruction frequencies, and structural stability can all be estimated from AI-derived trajectories and time-resolved measurements [22]. Such information provides direct insight into the physical mechanisms governing nanoparticle evolution and establishes quantitative links between microscopic observations and macroscopic materials behavior. Recent developments increasingly integrate detection, segmentation, denoising, and tracking into unified end-to-end workflows capable of transforming raw microscopy videos directly into physically meaningful descriptors [47, 99]. This progression represents an important step toward automated characterization systems capable of simultaneously observing, quantifying, and interpreting dynamic nanoscale phenomena.

### 7.4 Evaluation, Benchmarking, and Scientific Validation

Benchmarking spatiotemporal DL presents challenges fundamentally different from those encountered in static image analysis. Whereas detection and segmentation evaluate spatial localization and boundary agreement within individual images, dynamic microscopy requires simultaneous assessment of spatial accuracy, temporal consistency, object identity preservation, and the reliable identification of physically meaningful events occurring throughout image sequences. Early tracking applications largely adopted metrics developed within the computer vision community, including Multiple Object Tracking Accuracy (MOTA), Multiple Object Tracking Precision (MOTP), centroid displacement error, trajectory reconstruction accuracy, and related measures [100]. These metrics remain valuable for evaluating nanoparticle tracking and motion reconstruction, particularly in liquid-phase TEM and environmental microscopy where particle trajectories constitute primary experimental outputs [100].

However, particle tracking represents only the first level of scientific interpretation. In many nanoparticle systems, the principal objective is not determining where particles move but understanding how they transform during that motion. Processes including coalescence, sintering, defect formation, phase transitions, dissolution, and surface restructuring frequently require interpretation extending well beyond conventional trajectory analysis [22, 48]. Evaluating AI performance under these conditions becomes substantially more difficult because ground-truth annotations often depend on expert interpretation and may themselves contain ambiguity. When DL reveals transient structural states hidden beneath experimental noise, conventional image-quality metrics become insufficient for determining whether recovered features correspond to genuine physical phenomena or algorithm-induced artifacts. Increasing emphasis is therefore being placed on validation through consistency with established physical mechanisms, complementary experimental observations, and independent measurements rather than image similarity alone (see **Table 4**) [22, 100].

Table 4. Representative evaluation methodologies for AI-assisted dynamic and *in situ* nanoparticle microscopy.

| Method Family | Representative Application | Typical Metrics | Principal Limitation |
|---|---|---|---|
| Particle tracking | CNN-based tracking | MOTA, MOTP, trajectory error | Identity preservation |
| Dynamic defect analysis | Deep neural networks | Tracking accuracy, event detection | Annotation ambiguity |
| Liquid-phase nanoparticle dynamics | CNN + temporal association | Centroid error, displacement accuracy | Particle overlap |
| AI-assisted automated experimentation | Active learning / adaptive microscopy | Detection success rate, acquisition efficiency | Task-specific evaluation |
| Deep video denoising | Unsupervised temporal restoration | SNR improvement, FRC, temporal consistency | Structural validation |

As in structural inference, no universally accepted benchmarking framework currently exists for dynamic microscopy. Reported metrics vary substantially depending on the scientific objective, with tracking-focused studies emphasizing localization accuracy, restoration methods emphasizing temporal consistency and signal recovery, and mechanistic investigations increasingly evaluating recovery of physically meaningful quantities rather than image-level agreement. A particularly promising direction involves benchmarking directly against measurable physical outputs such as diffusion coefficients, growth rates, coalescence kinetics, activation energies, surface mobility, or phase-transition dynamics [95, 96]. Such quantities provide a natural bridge between computational performance and scientific utility, moving evaluation beyond purely algorithmic criteria toward metrics directly relevant to materials science. Similarly, uncertainty quantification is likely to become increasingly important as AI systems begin interpreting rare events, transient structural states, and highly dynamic nanoscale phenomena. Confidence estimation, probabilistic tracking, Bayesian inference, and physics-informed uncertainty analysis offer promising avenues for improving the reliability and interpretability of AI-assisted dynamic microscopy.

### 7.5 From Dynamic Observation to Mechanistic Understanding

The progression from static image interpretation to spatiotemporal learning represents the culmination of the evolution described throughout this review. Artificial intelligence is no longer limited to identifying particles, restoring images, or inferring hidden structures from individual projections. Instead, it increasingly enables direct quantitative analysis of evolving physical processes, transforming electron microscopy from a descriptive characterization technique into a framework for investigating nanoscale mechanisms and kinetics [48, 101]. Perhaps more importantly, the boundaries between image acquisition, image restoration, structural inference, tracking, and physical interpretation are becoming progressively blurred. Emerging AI pipelines increasingly integrate these components into unified analytical frameworks capable of converting raw microscopy data into mechanistic insight with minimal human intervention [47, 99, 102]. Recent developments further extend this concept toward intelligent microscopy ecosystems in which AI agents coordinate image acquisition, analysis, interpretation, and workflow management through event-driven architectures and adaptive decision making. Such approaches have the potential to integrate large language models, multimodal foundation models, and microscopy-specific AI into unified experimental environments capable of assisting both data acquisition and scientific interpretation [103]. This convergence represents one of the clearest indicators that electron microscopy is entering a new era in which AI functions not simply as an analysis tool but as an integral component of scientific discovery itself.

## 8. Conclusions and Future Perspectives

Artificial intelligence has transformed nanoparticle microscopy from a predominantly descriptive imaging technique into an increasingly quantitative and predictive scientific discipline. As demonstrated throughout this review, ML and DL methodologies now span virtually every stage of the microscopy workflow, from automated particle detection and segmentation to atomic-resolution restoration, three-dimensional structural inference, and the quantitative analysis of dynamic nanoscale processes. More importantly, their role has progressively evolved beyond workflow automation toward the extraction of physically meaningful information that is difficult or impossible to obtain through conventional image analysis alone.

A central theme emerging from the current literature is the progressive increase in the level of scientific inference enabled by AI. Early applications focused primarily on replacing repetitive manual tasks such as particle counting, localization, and segmentation with automated image-analysis pipelines capable of processing large microscopy datasets efficiently and reproducibly. More recent developments have shifted toward recovering structural information obscured by experimental limitations, inferring latent three-dimensional characteristics from two-dimensional projections, and quantifying complex dynamic processes observed during in situ experiments. This evolution reflects a broader conceptual transition in which artificial intelligence increasingly functions not simply as an image-processing tool but as an inference engine operating at the interface between experiment, simulation, and physical modeling.

At the same time, the maturity of AI methodologies varies considerably across different application domains. Detection and segmentation have reached a level of development that increasingly supports routine deployment in experimental workflows, benefiting from standardized evaluation metrics, relatively large, annotated datasets, and well-established architectures such as YOLO, U-Net, and nnU-Net. In contrast, atomic-resolution restoration, three-dimensional structural inference, and spatiotemporal analysis remain active research frontiers in which methodological innovation continues to outpace standardization. In these areas, questions regarding physical consistency, uncertainty quantification, reproducibility, and scientific validation remain at least as important as improvements in numerical performance.

Another recurring theme throughout this review is the growing importance of physics-informed learning and simulation-based methodologies. Because experimentally verified ground truth is frequently unavailable in atomic-resolution microscopy, many of the most successful AI frameworks rely on physically realistic simulations, prior knowledge of image formation mechanisms, or explicit incorporation of crystallographic constraints into the learning process. Future progress will likely depend increasingly on closer integration between machine learning algorithms, electron microscopy physics, atomistic simulations, and materials informatics rather than on architectural innovation alone.

The review also highlights a significant need for improved benchmarking and validation standards. While relatively mature evaluation protocols exist for detection and segmentation, no comparable consensus currently exists for denoising, structural inference, or dynamic microscopy. The establishment of shared datasets, realistic simulation frameworks, community benchmark challenges, and physically meaningful validation criteria will therefore be essential for enabling rigorous comparison between competing methodologies and accelerating their adoption by the broader microscopy community. Equally important will be the incorporation of uncertainty quantification and explainability into AI-assisted workflows, ensuring that scientific conclusions derived from machine learning models remain transparent, interpretable, and reproducible.

Looking forward, several technological developments appear particularly likely to shape the next generation of AI-assisted nanoparticle microscopy. Self-supervised learning offers the possibility of reducing dependence on expensive manual annotations by exploiting the enormous volumes of unlabeled microscopy data already being generated worldwide. Foundation models and vision transformers may provide increasingly generalizable representations capable of transferring knowledge across different materials systems and imaging modalities. Multimodal learning architectures integrating microscopy, spectroscopy, atomistic simulations, metadata, and theoretical calculations could enable substantially richer representations of nanoscale materials than image-based approaches alone. Similarly, autonomous microscopy and closed-loop experimental frameworks promise to transform AI from a passive analysis tool into an active component of experimental design and data acquisition. Emerging event-driven AI agents and adaptive microscopy workflows indicate a future in which acquisition, reconstruction, analysis, and experimental decision-making are integrated within unified intelligent microscopy platforms [47, 103].

Perhaps the most significant long-term implication extends beyond microscopy itself. The progression described throughout this review—from detection and segmentation through restoration, structural inference, and dynamic analysis—reflects a broader transition toward AI systems capable of extracting mechanistic understanding rather than merely recognizing visual patterns. Future developments are likely to focus increasingly on recovering physically meaningful quantities such as diffusion coefficients, growth kinetics, activation energies, surface mobility, defect energetics, and structure-property relationships directly from experimental observations. Together with multimodal foundation models, adaptive acquisition, reinforcement learning, and AI agents, these developments point toward autonomous microscopy platforms capable of performing closed-loop characterization and accelerating materials discovery [36, 72, 103]. In this context, AI becomes not simply a computational convenience but an integral component of scientific reasoning and materials discovery.

The emerging opportunities and principal challenges expected to shape this transition are summarized in **Table 5**.

Table 5. Emerging trends, current limitations, and future opportunities in AI-assisted nanoparticle microscopy.

| Area | Current Status | Main Limitation | Future Direction |
|---|---|---|---|
| Detection | Highly mature; YOLO-based methods routinely achieve high precision and recall | Sensitivity to highly crowded or low-contrast datasets | Foundation-model-based universal detectors |
| Segmentation | U-Net and nnU-Net provide state-of-the-art performance | Dependence on annotated datasets | Prompt-based and foundation-model segmentation (SAM, MedSAM) |
| Atomic-resolution denoising | Rapidly advancing through self-supervised and physics-informed methods | Risk of hallucinated atomic features | Uncertainty-aware and physics-constrained restoration |
| 2D-to-3D structural inference | Promising proof-of-concept demonstrations | Lack of experimental ground truth and benchmarking standards | Probabilistic reconstruction and uncertainty quantification |
| Dynamic / in situ microscopy | Growing ability to track and quantify nanoscale dynamics | Scarcity of annotated video datasets | Transformer-based temporal models and autonomous experimentation |
| Simulation-based training | Widely used due to limited experimental labels | Domain gap between simulations and experiments | Hybrid simulation–experiment training pipelines |
| Benchmarking | Mature only for detection and segmentation | Lack of common datasets and metrics | Community benchmark datasets and open challenges |
| Explainability | Increasingly recognized as essential | Black-box decision making | Explainable AI and physically interpretable latent representations |
| Self-supervised learning | Emerging research area | Limited validation in microscopy applications | Large-scale pretraining on microscopy repositories |
| Vision Transformers | Demonstrated success in image representation learning | High computational cost and data requirements | Hybrid CNN–Transformer architectures |
| Foundation Models | Early adoption in microscopy | Limited domain-specific adaptation | Universal microscopy foundation models |
| Multimodal Learning | Mostly experimental | Integration of heterogeneous datasets | Joint learning from microscopy, spectroscopy, simulation, and metadata |
| Autonomous Microscopy | Early demonstrations in STEM | Experimental integration complexity | Closed-loop AI-guided experimentation |
| Commercial Deployment | Emerging platforms such as SenseAI | Limited transparency of proprietary algorithms | Standardized validation and reproducible deployment |
| Materials Discovery | Increasing use for structure–property analysis | Weak integration with predictive modeling | AI-guided materials design and inverse materials engineering |

Collectively, the evidence reviewed here suggests that nanoparticle microscopy is entering a new era in which advances in instrumentation and advances in AI evolve synergistically rather than independently. The ultimate impact of AI will therefore not be measured solely by improvements in image quality or numerical accuracy, but by its ability to generate reliable physical insight, accelerate materials discovery, and enable scientific observations that would otherwise remain beyond the reach of conventional experimental methodologies.

## Appendix A.

The primary focus of this review is the application of AI to nanoparticle microscopy rather than the detailed development of ML algorithms themselves. Nevertheless, the rapid growth of AI-assisted microscopy has introduced a wide range of computational architectures originating from computer vision, medical imaging, and data science. Readers from materials science backgrounds may therefore benefit from a concise overview of the principal methodologies discussed throughout the manuscript.

The architectures summarized in this appendix were not originally developed for microscopy applications. Instead, most emerged from broader advances in image classification, object detection, segmentation, representation learning, and multimodal artificial intelligence. Their subsequent adaptation to microscopy has enabled increasingly sophisticated analysis of nanoparticle images, ranging from automated particle detection and segmentation to atomic-resolution interpretation, structural inference, and dynamic analysis.

Artificial intelligence architectures used in nanoparticle microscopy can be broadly divided into several families according to their underlying learning strategy. Convolutional neural networks (CNNs) form the foundation of modern image analysis and learn hierarchical image features directly from raw image data, making them particularly effective for classification, detection, and segmentation tasks. Detection architectures such as Faster R-CNN and YOLO extend CNN-based approaches to identify and localize individual particles within an image, while segmentation networks such as U-Net, Mask R-CNN, and nnU-Net perform pixel-level classification to delineate particle boundaries and extract quantitative morphological information.

More recent developments include transformer architectures, which employ self-attention mechanisms to capture long-range relationships across an image and have demonstrated strong performance in classification and segmentation tasks. Self-supervised learning approaches, including SimCLR, DINO, and MAE, seek to learn meaningful image representations without requiring extensive manual annotation, an important advantage for microscopy datasets where labeled examples are often scarce. Generative models, such as generative adversarial networks (GANs) and diffusion models, are increasingly being explored for image restoration, denoising, super-resolution, and data augmentation. Finally, foundation models and multimodal architectures, including CLIP, SAM, and MedSAM, represent a new generation of highly transferable AI systems capable of adapting to multiple downstream tasks with limited additional training. Together, these developments illustrate the progression of artificial intelligence from task-specific image analysis toward increasingly flexible, generalizable, and physically informed learning frameworks.

**Table A1** summarizes the principal artificial intelligence architectures that have influenced nanoparticle microscopy over the past decade. The table traces each methodology back to its original publication and application domain, illustrating how advances initially developed for computer vision and biomedical imaging were subsequently transferred to materials characterization. Together, these architectures form the computational foundation underlying most contemporary AI-assisted microscopy workflows.

Table A1. Representative AI architectures used in nanoparticle microscopy, their original applications, and adaptation to microscopy.

| **Category** | **Architecture** | **Year** | **Original Reference** | **DOI** | **Initial Application** | **Core Innovation** | **First Use in Microscopy** | **Typical Nanoparticle Microscopy Applications** |
|---|---|---|---|---|---|---|---|---|
| CNN | AlexNet | 2012 | Krizhevsky *et al.* [53] | 10.1145/3065386 | ImageNet image classification | Deep convolutional feature extraction | Microstructure and defect classification | Particle classification, morphology recognition |
| CNN | VGG-16 | 2014 | Simonyan & Zisserman [104] | 10.48550/arXiv.1409.1556 | Large-scale image classification | Deep sequential convolutional blocks | Microscopy feature extraction | Transfer learning, feature representation |
| CNN | ResNet | 2015 | He *et al.* [105] | 10.1109/CVPR.2016.90 | Image recognition | Residual connections | Atomic defect identification in STEM | Defect recognition, HRTEM classification |
| CNN | DenseNet | 2017 | Huang *et al.* [106] | 10.1109/CVPR.2017.243 | Image recognition | Dense feature reuse | Atomic structure characterization | Atomic-resolution image analysis |
| CNN | EfficientNet | 2019 | Tan & Le [107] | 10.48550/arXiv.1905.11946 | Efficient image classification | Compound model scaling | Transfer learning for microscopy | Classification with limited datasets |
| Detection | Faster R-CNN | 2015 | Ren *et al.* [71] | 10.1109/TPAMI.2016.2577031 | Generic object detection | Region Proposal Network (RPN) | Nanoparticle localization in TEM | Particle counting and detection |
| Detection | YOLO | 2016 | Redmon *et al.* [70] | 10.1109/CVPR.2016.91 | Real-time object detection | Single-stage detection | Automated nanoparticle detection | High-throughput TEM analysis |
| Detection | RetinaNet | 2017 | Lin *et al.*[108] | 10.1109/ICCV.2017.324 | Dense object detection | Focal loss | Small object detection in microscopy | Sparse particle detection |
| Detection | DETR | 2020 | Carion *et al.* [109] | 10.1007/978-3-030-58452-8_13 | End-to-end object detection | Transformer-based detection | Emerging microscopy detection studies | Detection in complex backgrounds |
| Segmentation | U-Net | 2015 | Ronneberger *et al.* [55] | 10.1007/978-3-319-24574-4_28 | Biomedical image segmentation | Encoder-decoder + skip connections | TEM nanoparticle segmentation | Particle morphology and sizing |

| Category | Architecture | Year | Original Reference | DOI | Initial Application | Core Innovation | First Use in Microscopy | Typical Nanoparticle Microscopy Applications |
|---|---|---|---|---|---|---|---|---|
| Segmentation | UNet++ | 2018 | Zhou *et al.* [110] | 10.1007/978-3-030-00889-5_1 | Medical image segmentation | Nested skip pathways | Microscopy segmentation | Complex particle boundaries |
| Segmentation | DeepLabv3+ | 2018 | Chen *et al.* [111] | 10.48550/arXiv.1802.02611 | Semantic segmentation | Atrous spatial pyramid pooling | Materials segmentation | Multi-scale particle analysis |
| Segmentation | Mask R-CNN | 2017 | He *et al.* [73] | 10.1109/ICCV.2017.322 | Instance segmentation | Pixel-level object masks | Agglomerated nanoparticle analysis | Overlapping particle segmentation |
| Segmentation | nnU-Net | 2019 | Isensee *et al.* [79] | 10.1038/s41592-020-01008-z | Automated medical segmentation | Self-configuring segmentation framework | General microscopy segmentation | Large-scale automated analysis |
| Transformer | Vision Transformer (ViT) | 2021 | Dosovitskiy *et al.* [112] | 10.48550/arXiv.2010.11929 | Image classification | Self-attention representation learning | HRTEM defect classification | Structure recognition |
| Transformer | DeiT | 2021 | Touvron *et al.* [113] | 10.48550/arXiv.2012.12877 | Efficient transformer training | Knowledge distillation for transformers | Microscopy classification studies | Feature extraction |
| Transformer | Swin Transformer | 2021 | Liu *et al.* [114] | 10.1109/ICCV48922.2021.00986 | Dense vision tasks | Hierarchical windowed attention | Microscopy segmentation | Multi-scale structural analysis |
| Transformer | Segmenter | 2021 | Sun *et al.* [115] | 10.1109/ICCV48922.2021.00719 | Semantic segmentation | Pure transformer segmentation | Experimental microscopy applications | Large-scale segmentation |
| Self-Supervised | SimCLR | 2020 | Chen *et al.* [116] | 10.48550/arXiv.2002.05709 | Representation learning without labels | Contrastive learning | Learning from microscopy archives | Unlabeled dataset analysis |
| Self-Supervised | DINO | 2021 | Caron *et al.* [117] | 10.48550/arXiv.2104.14294 | Self-supervised vision learning | Self-distillation | Microscopy representation learning | Feature discovery |
| Self-Supervised | MAE | 2021 | He *et al.* [118] | 10.48550/arXiv.2111.06377 | Masked image modeling | Reconstruction-based pretraining | Emerging microscopy applications | Foundation-model pretraining |
| Generative AI | GAN | 2014 | Goodfellow *et al.* [119] | 10.1145/3422622 | Generative modeling | Adversarial learning | Super-resolution microscopy | Image restoration, enhancement |
| Generative AI | CycleGAN | 2017 | Zhu *et al.* [120] | 10.1109/ICCV.2017.244 | Image-to-image translation | Unpaired domain translation | TEM/STEM enhancement | Contrast normalization |
| Generative AI | Diffusion Models | 2020 | Ho *et al.* [121] | 10.48550/arXiv.2006.11239 | Probabilistic image generation | Iterative denoising | Emerging restoration studies | Denoising and reconstruction |
| Foundation Model | CLIP | 2021 | Radford *et al.* [122] | 10.48550/arXiv.2103.00020 | Vision-language understanding | Multimodal representation learning | Materials informatics | Image–metadata integration |
| Foundation Model | SAM | 2023 | Kirillov *et al.* [75] | 10.48550/arXiv.2304.02643 | Universal image segmentation | Promptable segmentation | Electron microscopy segmentation | General-purpose segmentation |
| Foundation Model | MedSAM | 2024 | Ma *et al.* [76] | 10.1038/s41467-024-44824-z | Medical image segmentation | Domain-adapted SAM | Scientific image segmentation | TEM/HRTEM segmentation |
| Temporal Models | CNN-LSTM | ~2017 | Various foundational works | — | Video analysis | Spatial–temporal modeling | In situ TEM videos | Dynamic nanoparticle analysis |
| Reinforcement Learning | Deep Q-Network (DQN) | 2015 | Mnih *et al.* [123] | 10.1038/nature14236 | Sequential decision-making | Deep reinforcement learning | Autonomous microscopy | Adaptive experiment control |

Most architectures summarized in Table A1 were originally developed for computer vision, biomedical imaging, or ML applications unrelated to microscopy. Their inclusion reflects the methodologies that have subsequently been adapted to nanoparticle characterization workflows discussed throughout this review. The “First Use in Microscopy” column refers to the initial class of microscopy applications in which the architecture became established rather than a single definitive publication, as adoption often occurred simultaneously across multiple imaging domains.

A comparison of these architectures reveals a broader technological evolution extending beyond individual models. Early microscopy studies primarily focused on adapting CNNs for image classification, detection, and segmentation tasks [17, 18, 28, 31, 69]. As datasets increased in size and complexity, attention shifted toward physics-informed learning, automated feature extraction, and the integration of simulation-generated training data. More recently, transformer architectures, self-supervised learning, foundation models, and autonomous experimentation have emerged as key drivers of methodological innovation. **Table A2** summarizes these major developments and illustrates the progression of AI-assisted microscopy from image interpretation toward increasingly sophisticated forms of structural and physical inference.

Table A2. Evolution of AI methodologies in nanoparticle microscopy.

| Period | Dominant Architectures | Main Objective | Representative Developments |
|---|---|---|---|
| 2015–2017 | CNNs, U-Net | Automated image interpretation | Classification and segmentation |
| 2017–2020 | ResNet, DenseNet, Faster R-CNN | Feature learning and detection | Defect recognition, particle detection |
| 2020–2022 | YOLOv4, nnU-Net, SimCLR | High-throughput characterization | Automated analysis pipelines |
| 2021–2023 | Physics-informed CNNs, ViT | Atomic-resolution interpretation | HRTEM and STEM analysis |
| 2022–2024 | Vision Transformers, Swin | Long-range feature modeling | Multi-scale microscopy analysis |
| 2023–2025 | SAM, MedSAM | Generalizable segmentation | Foundation-model adoption |
| 2024–Present | Multimodal AI, Autonomous Microscopy | Scientific inference | Closed-loop experimentation |

While Table A2 highlights the chronological evolution of AI methodologies, researchers typically select architectures according to the scientific problem being addressed rather than their historical development. Consequently, multiple AI approaches frequently coexist within a single microscopy workflow, with some optimized for particle detection and segmentation, others for image restoration and structural reconstruction, and still others for dynamic analysis of time-resolved experiments. **Table A3** therefore reorganizes the methodologies discussed throughout this review according to their primary role in nanoparticle microscopy and provides an overview of the current maturity of each application area.

Table A3. Mapping AI architectures to nanoparticle microscopy tasks.

| Task | Dominant Architectures | Representative Studies | Current Maturity |
|---|---|---|---|
| Particle detection | YOLO, Faster R-CNN | DetectNano | Mature |
| Particle segmentation | U-Net, nnU-Net, SAM | Madsen *et al.*, nNPipe | Mature |
| Atomic defect recognition | ResNet, DenseNet, ViT | Ziatdinov *et al.* | Developing |
| Atomic-resolution denoising | U-Net variants, self-supervised CNNs | AI4AM, Crozier *et al.* | Rapidly advancing |
| 2D-to-3D structural inference | CNNs, Transformers | AI4AM and related studies | Emerging |
| In situ TEM analysis | CNN-LSTM, Transformers | Yao *et al.*, Ziatdinov *et al.* | Emerging |
| Autonomous microscopy | Active learning, reinforcement learning | Automated STEM studies | Early stage |

Several important trends emerge from Tables A1–A3. First, the field has evolved from supervised convolutional architectures designed primarily for image interpretation toward increasingly general learning frameworks capable of handling complex structural and temporal information. Second, the distinction between image analysis and scientific inference is becoming progressively less pronounced, particularly in areas such as atomic-resolution reconstruction, dynamic microscopy, and autonomous experimentation. Finally, recent developments in foundation models, multimodal learning, and self-supervised approaches suggest that future AI systems will operate not only as image-processing tools but also as active components of scientific discovery. The technological progression of these methodologies and their impact on nanoparticle microscopy is summarized schematically in **Figure A1**.

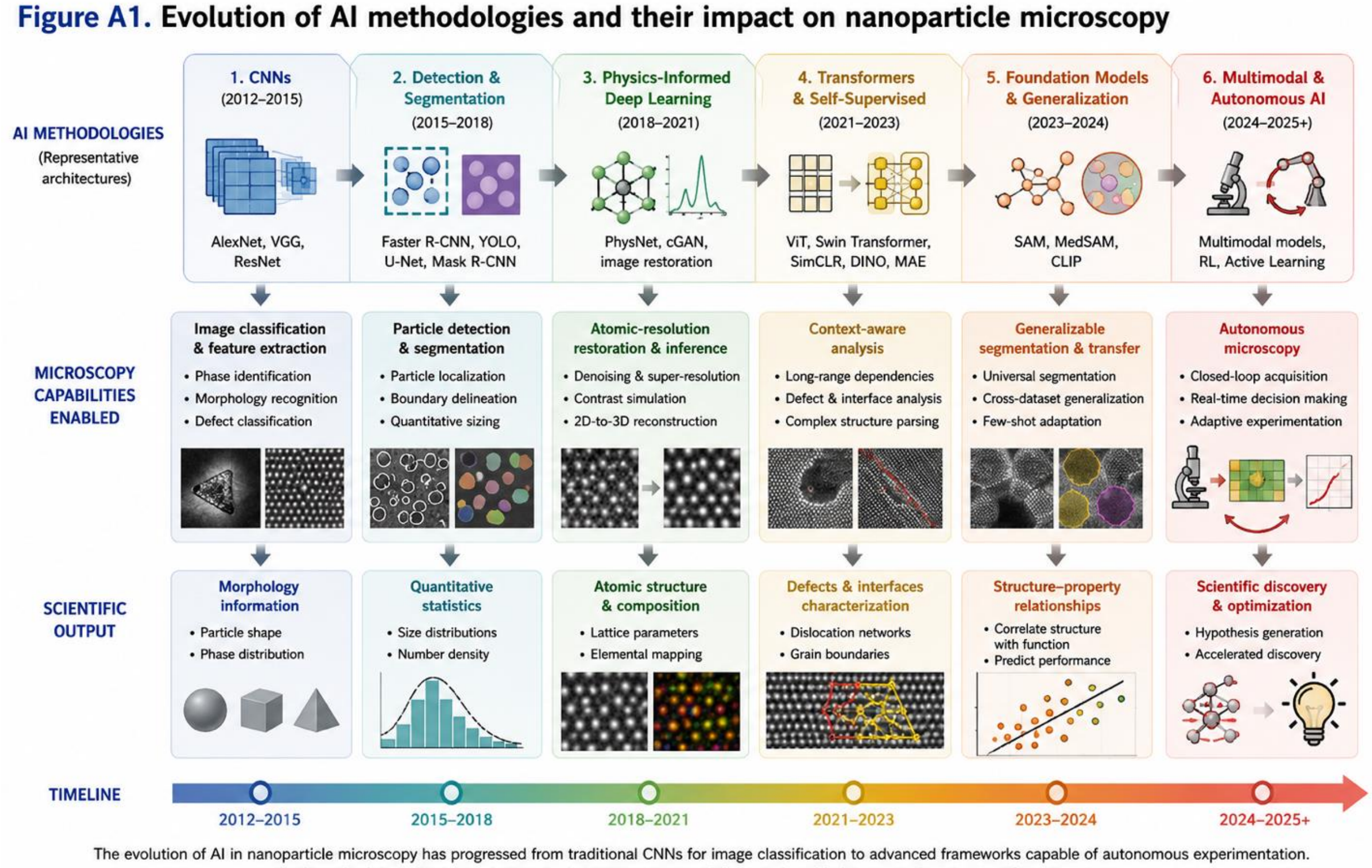

Figure A1. Evolution of artificial intelligence methodologies and their impact on nanoparticle microscopy. The field has progressed from convolutional neural networks primarily used for image classification and feature extraction toward increasingly sophisticated frameworks capable of performing particle detection, segmentation, atomic-resolution reconstruction, and dynamic analysis. Recent developments in transformer architectures, foundation models, and autonomous learning systems are shifting the role of AI from image interpretation toward scientific inference and closed-loop experimentation, enabling the extraction of increasingly complex structural and functional information from microscopy data.

The architectural and methodological developments summarized in this appendix provide the computational foundation for the application-oriented discussions presented throughout the main text. However, the evolution of AI-assisted nanoparticle microscopy has not been driven by isolated algorithms but by a continuous interplay between advances in computer vision, ML, electron microscopy, and materials science. The representative publications compiled in **Table A4** therefore complement the preceding summaries by highlighting influential studies across the major themes of this review, including detection and segmentation, atomic-resolution restoration, three-dimensional structural inference, and spatiotemporal analysis. Rather than serving as an exhaustive bibliography, the table is intended to provide readers with a practical roadmap to the key methodological milestones and representative contributions that have shaped the rapid development of the field.

Table A4. Representative studies in AI-assisted nanoparticle microscopy organized by primary methodological focus and application domain.
Sections:

- 3 = ML/DL Methods (general / foundational / reviews)
- 4 = Detection & Segmentation (TEM / morphology)
- 5 = Atomic Resolution Denoising & Restoration (HRTEM/STEM)
- 6 = 2D → 3D Inference & Reconstruction
- **7** = *In Situ* / Spatiotemporal / Dynamics

| Year | Paper | DOI | Main Focus | Section(s) |
|---|---|---|---|---|
| 2017 | Ziatdinov *et al.*, Deep Learning of Atomically Resolved STEM Images [18] | 10.1021/acsnano.7b07504 | Atomic identification & tracking | 3,5,7 |
| 2018 | Maksov *et al.*, Deep Learning Analysis of Defects in STEM [124] | 10.1038/s41524-019-0152-9 | Defect detection in STEM | 5 |
| 2019 | Oviedo *et al.*, Fast and interpretable classification of small XRD datasets [125] | 10.1038/s41524-019-0196-x | Early ML structural classification | 3 |
| 2020 | Lee *et al.*, Statistical Morphology via ML [17] | 10.1021/acsnano.0c06809 | Morphology statistics | 3,4 |
| 2020 | Zhu *et al.*, Lattice Spacing via DL [126] | 10.1007/s40843-020-1368-7 | HRTEM lattice extraction | 5 |
| 2020 | Horwath *et al.*, DL features in HRTEM segmentation [127] | 10.1038/s41524-020-00363-x | Interpretability in HRTEM | 5 |
| 2020 | Ede, Exit wave DL [128] | 10.5281/ZENODO.4399748 | Physics-informed reconstruction | 5,6 |
| 2021 | Koyama *et al.*, CNN analysis of metallic nanoparticles [129] | 10.1016/j.jmmm.2021.168225 | TEM classification | 3,4 |
| 2021 | Shen *et al.*, In situ defect DL framework [130] | 10.1016/j.commatsci.2021.110560 | Temporal defect tracking | 7 |
| 2021 | Kashin *et al.*, Neural network video EM analysis [131] | 10.1002/smll.202007726 | Video-based dynamics | 7 |
| 2021 | Liu *et al.*, Boundary-attention network for catalyst particles [132] | 10.1016/j.mtcomm.2021.102728 | Geometry extraction | 4 |
| 2021 | Akers *et al.*, Few-shot EM segmentation [133] | 10.1038/s41524-021-00652-z | Low-data segmentation | 3,4 |
| 2022 | Sun *et al.*, Morphology analysis framework [134] | 10.1039/d2nr01029a | SEM/TEM morphology | 4 |
| 2022 | Lee *et al.*, STEM $MoS_2$ defect classification [135] | 10.1021/acs.nanolett.2c00550 | Atomic defect inference | 5,6 |
| 2022 | Cha *et al*., Low-dose STEM tomography via DL [136] | 10.1021/acsnano.2c00168 | Sparse-view reconstruction | 6 |
| 2022 | Ragone *et al.*, Element mapping in HEAs via DL [137] | 10.1016/j.commatsci.2021.110905 | STEM elemental inference | 5 |
| 2022 | Botifoll *et al.*, ML in electron microscopy review [28] | 10.1039/d2nh00377e | Review | 3 |
| 2022 | Cheng *et al.*, In situ TEM + ML review [138] | 10.1088/1674-4926/43/8/081001 | Review | 3,7 |
| 2023 | Yao L. Chen Q, ML in EM data analysis [80] | 10.1016/B978-0-323-85796-3.00010-X | Overview | 3 |
| 2023 | Treder *et al.*, nNPipe catalyst morphology pipeline [139] | 10.1038/s41524-022-00949-7 | Catalyst segmentation | 4 |
| 2023 | Kulesh *et al.*, DL segmentation optimization [140] | 10.1016/j.actamat.2023.119039 | Data-driven TEM analysis | 4 |
| 2023 | Larsen *et al.*, Exit wave reconstruction ML [81] | 10.1016/j.ultramic.2022.113641 | Physics reconstruction | 5,6 |
| 2023 | Larsen *et al.*, Noise limits in HRTEM DL segmentation [82] | 10.1016/j.ultramic.2023.113803 | Reliability limits | 5 |

| Year | Paper | DOI | Main Focus | Section(s) |
|---|---|---|---|---|
| 2023 | Zhu *et al.*, DL-assisted HRTEM nanoparticle analysis [141] | 10.1039/d3nr03061j | HRTEM NP analysis | 5,6 |
| 2023 | Kalinin *et al.*, ML for automated STEM experimentation [47] | 10.1038/s41524-023-01142-0 | Closed-loop microscopy | 7 |
| 2023 | Ragone *et al.*, DL modeling in microscopy review [142] | 10.1016/j.pmatsci.2023.101165 | Review | 3 |
| 2023 | Cheng *et al.*, Liquid-phase TEM single-molecule DL [143] | 10.1039/d2cc05354c | Dynamics in liquid TEM | 7 |
| 2023 | Azimi *et al.*, Deep learning microstructure classification [144] | 10.1038/s41598-018-20037-5 | Classification | 3,4 |
| 2024 | Wang *et al.*, DDDNet segmentation model [145] | 10.1063/5.0228023 | TEM segmentation | 4 |
| 2024 | Moreira *et al.*, Ag nanoparticle segregation via ML [146] | 10.1021/acsanm.3c05495 | STEM compositional inference | 5,6 |
| 2024 | Sun *et al.*, Liquid-phase TEM DL quantitative analysis [56] | 10.1039/d3nr04480g | Dynamic NP analysis | 7 |
| 2024 | Chen *et al.*, Advancing EM using DL review [57] | 10.1088/2515-7639/ad229b | Review | 3 |
| 2024 | Fan *et al.*, Self-supervised microstructure learning [58] | 10.1016/j.mattod.2024.05.003 | SSL features | 3 |
| 2024 | Chen *et al.,* Vision transformers for microstructure [59] | 10.1039/D3DD00198A | Review | 3 |
| 2024 | Zhou *et al.*, Multi-scale microstructure DL [60] | 10.1038/s41524-023-01011-y | Multi-scale learning | 3 |
| 2024 | Zhou *et al.*, TEM nanoparticle segmentation [61] | 10.3390/nano14141169 | NP segmentation | 4 |
| 2024 | Faraz *et al.*, TEM nanoparticle detection ETEM tracking [62] | 10.1038/s41598-022-06308-2 | Tracking | 7 |
| 2024 | Lu *et al.*, ML for M2D materials [147] | 10.1002/advs.202305277. | Review | 1 |
| 2025 | Ferji K., DetectNano | 10.1039/D5NR02446C | YOLOv8 nanoparticle detection | 4 |
| 2025 | Giner *et al.*, HRTEM 3D inference [148] | — | Shape + denoising | 5,6 |
| 2025 | Kazimi *et al.,* Semantic segmentation HRTEM improvements [63] | 10.1093/mam/ozae093 | HRTEM segmentation | 5 |
| 2025 | Day *et al.*, Nanoparticle segmentation automation [64] | 10.1038/s41598-025-01337-z | Pipeline | 4 |
| 2025 | Kim *et al.*, Self-supervised LPTEM analysis [65] | 10.1126/sciadv.ads5552 | Dynamic learning | 7 |
| 2025 | Van *et al.*, Foundation models in materials imaging [36] | arXiv:2506.20743v1 | Large models | 3 |
| 2025 | Gao *et al.,* AI polymer nanoparticle design [66] | 10.1002/adma.202516857 | Application | 3 |
| 2026 | Han *et al.*, DL framework for nano-segmentation [149] | 10.55092/aimat20260006 | Segmentation | 3 |
| 2024–2026 | NanoScience Instruments SenseAI | https://www.nanoscience.com/products/senseai-software/ | Commercial AI-powered microscopy analysis platform; automated particle identification and characterization | 3,4 |

## Acknowledgments

PG was supported by a Beatriz Galindo Senior Fellowship (BG23/00144). Work performed at GTIIT was supported by funding from the Guangdong Technion Israel Institute of Technology.

## Conflicts of Interest

The authors declare no conflicts of interest

## Data Availability Statement

The data that support the findings of this study are available from the corresponding author upon reasonable request.